\documentclass[prb,twocolumn]{revtex4}
\usepackage{times}
\usepackage[dvips]{graphicx}
\usepackage{latexsym,amsmath,amssymb,bm,euscript}
\usepackage[dvips]{color}

\def\ra{\rangle} 
\def\la{\langle} 
\def\up{\uparrow}
\def\dn{\downarrow}
\def\Hc{{\rm H.c.}}
\def\pd{{ \phantom\dagger}}

\begin{document}

\title{Strong-Coupling Phases of Frustrated Bosons on a 2-leg Ladder 
with Ring Exchange}

\author{D. N. Sheng$^1$, Olexei I. Motrunich$^2$,  Simon Trebst$^3$, 
Emanuel Gull$^4$, Matthew P. A. Fisher$^3$}
\address{
$^1$Department of Physics and Astronomy, California State
University, Northridge, California 91330\\
$^2$Department of Physics, California Institute of Technology,
Pasadena, CA 91125\\
$^3$Microsoft Research, Station Q, University of California, 
Santa Barbara, California 93106\\
$^4$Theoretische Physik, Eidgen\"ossische Technische Hochschule Z\"urich, 
CH-8093 Z\"urich, Switzerland}

\date{\today}

\begin{abstract}

Developing a theoretical framework to access quantum phases of 
itinerant bosons or fermions in two dimensions (2D) that exhibit 
singular structure along surfaces in momentum space but have no
quasi-particle description remains as a central challenge in the 
field of strongly correlated physics.  In this paper we propose that 
distinctive signatures of such 2D strongly correlated phases will be 
manifest in quasi-one-dimensional ``$N$-leg ladder" systems.  
Characteristic of each parent 2D quantum liquid would be a precise 
pattern of 1D gapless modes on the $N$-leg ladder.
These signatures could be potentially exploited to approach the 2D phases
from controlled numerical and analytical studies in quasi-1D.
As a first step we explore itinerant boson models with a frustrating 
ring exchange interaction on the 2-leg ladder, searching for signatures 
of the recently proposed two-dimensional d-wave correlated Bose liquid 
(DBL) phase.  A combination of exact diagonalization, density matrix 
renormalization group, variational Monte Carlo, and bosonization 
analysis of a quasi-1D gauge theory, all provide compelling evidence 
for the existence of a new strong-coupling phase of bosons on the 
2-leg ladder which can be understood as a descendant of the 
two-dimensional DBL.  We suggest several generalizations to quantum spin 
and electron Hamiltonians on ladders which could likewise reveal
fingerprints of such 2D non-Fermi liquid phases.

\end{abstract}

\maketitle

\section{Introduction}
\label{sec:intro}

During the past decade it has become abundantly clear that
there are several sub-classes of two-dimensional (2D) spin liquids.
\cite{LeeNagaosaWen}
Perhaps best understood are the topological spin liquids that
support gapped excitations which carry fractional quantum numbers.
\cite{RSSpN, Wen, topth, MoeSon}
Another possibility are gapless spin liquids with no topological structure.
These gapless spin liquids will generically exhibit spin correlations that 
decay as a power law in space and which can oscillate at particular 
wavevectors.  In such ``algebraic" or ``critical" spin liquids
\cite{WenPSG, Rantner, Hermele, LeeNagaosaWen} these wavevectors are 
limited to a finite discrete set, and their effective field theories 
will often exhibit a relativistic structure -- as is the case when the 
spinons in a slave particle construction are massless Dirac fermions 
coupled to a U(1) gauge field.  
However, it is possible that the spin correlations exhibit singularities
along surfaces in momentum space.  These are analogous to the Fermi 
surfaces in a Fermi liquid but in such ``quantum spin-metals" a 
quasiparticle picture will presumably be inapplicable.
Spin liquids with a spinon Fermi surface are examples of particular 
interest, and have been studied in a number of RVB works
\cite{Baskaran, IoffeLarkin, LeeNagaosa, Polchinski, Altshuler, LeeNagaosaWen} 
-- most recently as a candidate for a spin liquid phase
\cite{Shimizu03, ringxch, SSLee}
in the organic compound $\kappa$-(ET)$_2$Cu$_2$(CN)$_3$.

\begin{figure}
\centerline{\includegraphics[width=\columnwidth]{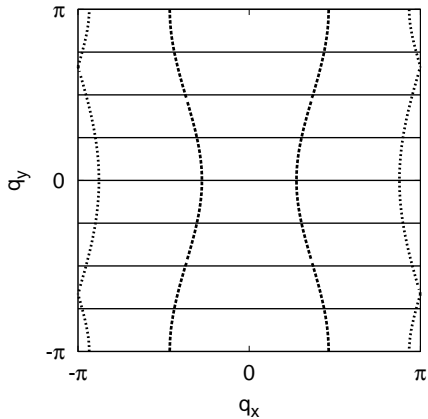}}
\vskip -2mm
\caption{Schematics of discrete $q_y = 2\pi/N$ cuts through Bose surfaces
in some anisotropic 2D system placed on an $N$-leg ladder; 
here for illustration $N=8$.  
Such Bose surfaces appear in a DBL construction described in the text; 
to avoid clutter, we do not show the Fermi surfaces of the $d_1$ and 
$d_2$ slave particles.
}
\label{fig:qycuts}
\end{figure}

Developing a theoretical framework for itinerant 2D non-Fermi liquids is 
arguably even more challenging than for spin liquids.
Towards this end recent work\cite{DBL} explored the possibility of an
uncondensed quantum phase of itinerant bosons in two dimensions which is 
a conducting fluid but not superfluid.   Because of the characteristic
nodal structure observed by moving one boson around another,
Ref.~\onlinecite{DBL} called such phases 
``d-wave Correlated Bose Liquids'' (DBL).
Various physical correlators were found to be singular across surfaces 
in momentum space in the DBL.  For example, the boson occupation number 
$n({\bf q})$ is singular at some ``Bose surfaces'' like those 
illustrated in Fig.~\ref{fig:qycuts}.  
Other examples with critical surfaces have been studied recently in 
Ref.~\onlinecite{Senthil}.

The central idea underlying this paper is that 2D phases of quantum 
spins and itinerant fermions or bosons with singular surfaces in 
momentum space should have definite signatures when restricted to a 
quasi one-dimensional (1D) geometry, e.g.\ when the system is placed 
onto an $N$-leg ladder.
To be concrete, suppose we have a singular 2D surface in momentum space
in the ground state of some square lattice quantum Hamiltonian.
If we put the Hamiltonian on an $N$-leg ladder, the transverse momentum
$q_y$ becomes discrete.  With large $N$ one would expect that the 
short-range energetics that stabilizes the 2D phase will still be 
present on the ladder and will lead to a set of 1D gapless modes 
where the discrete momenta cut through the singular surface, 
as illustrated in Fig.~\ref{fig:qycuts}. 

The long wavelength physics of these 1D modes will be described 
in terms of some multi-channel Luttinger liquid, behavior different 
than the parent 2D state.
But the number of gapless 1D modes and their respective momenta
characterizing the oscillatory power law decays would provide a 
distinctive ``fingerprint" of the parent 2D liquid.
For example, the number of gapless 1D modes would grow linearly with $N$ 
and the sum of their momenta could satisfy some generalized Luttinger 
volume.  Moreover, these signatures could be potentially exploited to 
approach the 2D phases from controlled numerical and analytical studies 
in quasi-1D.
Indeed, as we demonstrate explicitly in this paper, the (putative) 
singular surface of the 2D DBL phase is already manifest on the 
2-leg ladder!

The itinerant uncondensed DBL of Ref.~\onlinecite{DBL} is constructed 
by writing a boson in terms of fermionic partons,
\begin{equation}
b^\dagger = d_1^\dagger d_2^\dagger ~,
\label{d1d2}
\end{equation}
and considering un-coupled mean field states of the $d_1$ and $d_2$.
The DBL phase on a square lattice is obtained by taking the 
$d_1$ fermions to hop preferentially along the $\hat{x}$-axis and the 
$d_2$ fermions to hop preferentially along the $\hat{y}$-axis.
One can enforce the square lattice symmetry in the boson state
by requiring that the $d_1$ and $d_2$ hopping patterns are related by a 
90 degree rotation.
To recover the physical Hilbert space requires the condition
$d_1^\dagger d^\pd_1 = d_2^\dagger d^\pd_2 = b^\dagger b$ 
at each site.
Within a Gutzwiller projection approach the corresponding boson 
wavefunction is taken as product of two fermion Slater determinants,
\begin{equation}
\Psi_{\rm bos}(r_1, r_2, \dots) = \Psi_{d_1}(r_1, r_2, \dots) \cdot
\Psi_{d_2}(r_1, r_2, \dots) ~.
\label{PsiDBL}
\end{equation}
In the DBL the $d_1$ and $d_2$ Fermi surfaces are different,
compressed in the $\hat{x}$ and $\hat{y}$-directions, respectively.
This results in a characteristic d-wave-like nodal structure observed 
when moving one boson around another in $\Psi_{\rm bos}$.
The boson momentum distribution function $n({\bf q})$ in the DBL is 
singular on two surfaces that are constructed from the $d_1$ and $d_2$ 
Fermi surfaces as enveloping ${\bm k}_{F1} \pm {\bm k}_{F2}$ surfaces.
More details are in Ref.~\onlinecite{DBL}, while Fig.~\ref{fig:qycuts} 
shows one example of the ${\bm k}_{F1} \pm {\bm k}_{F2}$ loci for
an open $d_1$ Fermi surface and a closed $d_2$ Fermi surface 
(the Fermi surfaces themselves are not shown; 
this DBL state would be anisotropic on the square lattice but is 
closer to the ladder states considered here where there is no 
90 degree rotation symmetry).
As an alternative to the Gutzwiller construction one can project into 
the physical Hilbert space within a gauge theory approach, coupling 
$d_1$ and $d_2$ with opposite gauge charges to an emergent 
U(1) gauge field.

Ref.~\onlinecite{DBL} also proposed a frustrated boson model with 
ring exchanges that can potentially realize such DBL phases:
\begin{eqnarray}
\label{Hring}
H &=& H_J + H_{\rm ring} ~,\\
H_J &=& -J \sum_{{\bf r};\, \hat{\mu} = \hat{x}, \hat{y}}
(b^\dagger_{\bf r} b^\pd_{{\bf r} + \hat{\mu}} + \Hc) ~,\\
H_{\rm ring} &=& K \sum_{\bf r} 
(b^\dagger_{\bf r} b^\pd_{{\bf r} + \hat{x}}
b^\dagger_{{\bf r} + \hat{x} + \hat{y}} b^\pd_{{\bf r} + \hat{y}}
+ \Hc) ~,
\end{eqnarray}
with $J, K > 0$.  
In addition to the usual boson hopping term this Hamiltonian
contains a four-site ring term acting on each square plaquette.
With $K$ positive this term is ``frustrating", violating the 
Marshall sign rule of the hopping term -- making the system 
intractable by quantum Monte Carlo simulations.  This boson Hamiltonian 
was constructed\cite{DBL} by taking the strong coupling limit of the 
lattice gauge theory for the $d_1$ and $d_2$ fermions.
Increasing the disparity in the fermion hopping along the $\hat{x}$ and 
$\hat{y}$ axes corresponds to an increase in $K$.

In this paper, we make a first step in the proposed program of ladder 
studies by exploring the ring model Eq.~(\ref{Hring}) on a 2-leg ladder.
Of course, the picture of discrete $q_y$ cuts through two-dimensional 
surfaces is pushed to the extreme here.
Nevertheless and quite remarkably, in a wide parameter regime of 
intermediate to large ring couplings, we find an unusual phase which 
can be understood via the DBL construction Eq.~(\ref{PsiDBL}).
This is a strong-coupling phase with no perturbative picture in
terms of the original bosons.  The slave-particle approach provides 
a new starting point, and the resulting gauge theory of the DBL can be 
solved by the conventional 1D bosonization tools,
\cite{KimLee, Hosotani, Mudry, Ivanov}
providing a consistent picture of the new phase.

The paper is organized as follows.  
Section~\ref{sec:DBL_2leg} sets the stage by describing DBL states 
on the 2-leg ladder.
The main Section~\ref{sec:JK} presents results of exact numerical 
studies of the $J-K$ model.  The focus here is on the DBL phase
that emerges when the conventional ``${\bf q} = {\bf 0}$ liquid'' is 
destroyed by the ring exchanges.  However, the phase diagram is 
even more rich and contains two more unusual phases in which 
bosons are paired; one of these phases can be accessed analytically 
as an instability of yet a different DBL phase.
Appendix~\ref{app:GT} summarizes the technical gauge theory description 
of the DBL phases while Appendix~\ref{app:oddends} presents simple 
analytical results for the ring model in some limiting cases.
Section~\ref{sec:concl} concludes with an outline of future 
program of ladder studies.

\section{DBL States for Hard-Core Bosons on a 2-Leg Ladder}
\label{sec:DBL_2leg}

\begin{figure}
\centerline{\includegraphics[width=\columnwidth]{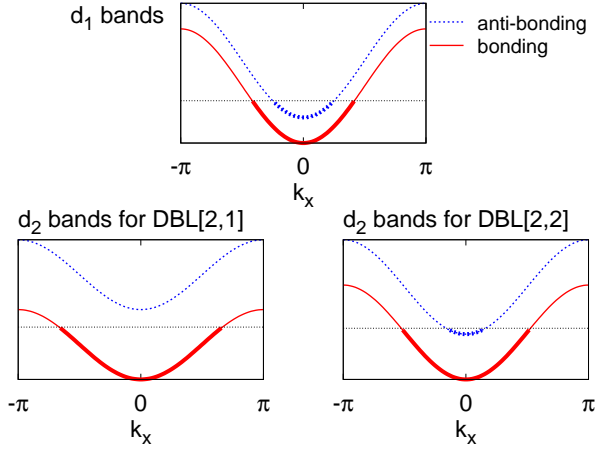}}
\vskip -2mm
\caption{(color online)
$d_1$ and $d_2$ bands for the DBL construction on the 2-leg ladder.  
Top panel: $d_1$ fermions partially occupy both the bonding and 
anti-bonding bands.
Bottom left:  In the DBL[2,1] phase, $d_2$ fermions occupy only
the bonding band.
Bottom right:  In the DBL[2,2] phase, $d_2$ fermions partially occupy 
both bands; this state is likely unstable towards a boson-paired phase, 
as discussed in Sec.~\ref{subapp:DBL22instab}.
In the wavefunction construction, only the occupied orbitals and not 
the band dispersions are important.
}
\label{fig:bands}
\end{figure}

The DBL construction Eq.~(\ref{PsiDBL}) proceeds by taking distinct 
hopping ground states for the $d_1$ and $d_2$ fermions.  
The hopping patterns of the two fermion species can be independent 
and need to only respect the ladder symmetries.  There is a bonding 
and an antibonding band for each species.  
We take the $d_1$ fermions to hop more strongly along the chains 
and always assume that the chemical potential crosses both bands as 
illustrated in Fig.~\ref{fig:bands}.
We take the $d_2$ fermions to hop more strongly on the rungs and
consider two cases shown in the bottom panels in Fig.~\ref{fig:bands}: 
The ``DBL[2,1] phase" when only the bonding band of the $d_2$ fermions
crosses the Fermi level, and the ``DBL[2,2] phase" when both $d_2$ bands 
are partially occupied
-- the first and second argument in the square brackets refer to the 
number of involved $d_1$ and $d_2$ bands respectively.

Let us first discuss properties of the DBL[2,1] phase.
We denote the Fermi momenta of the $d_1$ bonding and antibonding bands 
as $k_{F1}^{(0)}$ and $k_{F1}^{(\pi)}$, corresponding to the transverse
momenta $k_y=0$ and $k_y=\pi$ respectively (which is a convenient 
alternative to the quantum number for the leg interchange symmetry).
We denote the Fermi momentum of the $d_2$ bonding band as $k_{F2}$ 
(there is no $d_2$ antibonding occupation in the DBL[2,1]).  
The Fermi momenta satisfy 
$k_{F1}^{(0)} + k_{F1}^{(\pi)} = k_{F2} = 2 \pi \rho$, where $\rho$ is 
the boson density per site and remembering that there are two sites in 
each rung while the lattice spacing is set to one.

In a mean field treatment, the boson Green's function is a product of
two fermion Green's functions and therefore oscillates at different
possible $k_{F1} \pm k_{F2}$ wavevectors and decays as $1/x^2$.
Going beyond the mean field as described in Appendix~\ref{app:GT},
we expect that the oscillations at wavevectors
$q_x^{(0/\pi)} = k_{F2} - k_{F1}^{(0/\pi)}$ have slower power law decay 
while the other two wavevectors $k_{F2} + k_{F1}^{(0/\pi)}$ have faster 
power law decay than the mean field.  The complete result is, 
dropping all amplitudes,
\begin{eqnarray}
G_b({\bf r}) \sim 
\frac{\cos[(k^\pd_{F2} - k_{F1}^{(0)}) x]}{x^{2 - \eta^{(0)}}}
+ e^{i\pi y} \frac{\cos[(k^\pd_{F2} - k_{F1}^{(\pi)}) x]}
                  {x^{2 - \eta^{(\pi)}}}\\
+ \frac{\cos[(k^\pd_{F2} + k_{F1}^{(0)}) x]}{x^{2 + \dots}}
+ e^{i\pi y} \frac{\cos[(k^\pd_{F2} + k_{F1}^{(\pi)}) x]}{x^{2 + \dots}}.
\end{eqnarray}
The first line shows the enhanced correlators.  These can be guessed 
using an ``Amperean rule'' mnemonic inherited from the fermion-gauge 
studies in 2D and verified by the bosonization solution in 
Appendix~\ref{app:GT}:  Composites that involve fermion bi-linears with 
opposite group velocities but that produce parallel gauge currents are 
enhanced by the gauge fluctuations.
Thus, low-energy ``bosons'' $d_{1L}^{(0/\pi) \dagger} d_{2R}^\dagger$ 
carrying $q_x^{(0/\pi)}$ are expected to be enhanced.
The contributions in the second line, on the other hand,
can be shown to be always suppressed beyond the mean field.
Appendix~\ref{app:GT} presents a complete low-energy theory of the 
DBL[2,1] which -- once the gauge fluctuations are treated -- contains 
two phonon modes and in principle allows calculating all exponents.  
Since the power laws are affected also by short-range interactions, 
and there are many parameters allowed in the theory, we do not pursue 
this calculation in general, but mainly rely on the Amperean rules 
that capture the gauge fluctuation effects; 
this is reasonable if the gauge interactions dominate.

The boson singularities at non-zero wavevectors become manifest upon 
Fourier transformation of $G_b({\bf r})$ to obtain the boson occupation 
number $n({\bf q})$.  For example, at the enhanced wavevectors 
$(q_x^{(0)}, 0)$ and $(q_x^{(\pi)}, \pi)$, we have
\begin{equation}
n(q_x^{(k_y)} + \delta q_x, k_y) \sim |\delta q_x|^{1-\eta^{(k_y)}} ~.
\end{equation}

The boson density correlations are analyzed similarly.
In a naive mean field, 
$D_b({\bf r}) = \la n({\bf r}) n({\bf 0}) \ra \sim
D_{d1}({\bf r}) + D_{d2}({\bf r})$, which oscillates at various 
``$2k_F$'' vectors and decays as $1/x^2$.  
Beyond the mean field, the power laws get modified:
\begin{eqnarray}
&&D_{d1}({\bf r}) \sim \frac{1}{x^2} 
+ \frac{\cos[2 k_{F1}^{(0)} x]}{x^{2 - \dots}}
+ \frac{\cos[2 k_{F1}^{(\pi)} x]}{x^{2 - \dots}} \\
&&
+ e^{i\pi y}\frac{\cos[(k_{F1}^{(0)} + k_{F1}^{(\pi)}) x]}{x^{2 - \dots}}
+ e^{i\pi y}\frac{\cos[(k_{F1}^{(0)} - k_{F1}^{(\pi)}) x]}{x^{2 + \dots}}
~, \nonumber \\
&&D_{d2}({\bf r}) \sim \frac{1}{x^2} 
+ \frac{\cos[2 k_{F2} x]}{x^{2 - \dots}} ~.
\end{eqnarray}
The distinct momenta are: 
$(2 k_{F1}^{(0)}, 0)$, $(2 k_{F1}^{(\pi)}, 0)$, $(2 k_{F2}, 0)$,
$(k_{F1}^{(0)} + k_{F1}^{(\pi)}, \pi)$,
$(k_{F1}^{(0)} - k_{F1}^{(\pi)}, \pi)$, and $(0,0)$.
The first four involve a particle and a hole of the same species 
but with opposite group velocities and are therefore expected to be
enhanced by the gauge fluctuations; in the above equations, this is 
indicated schematically with $1/x^{2 - \dots}$.  
However, these correlators can also be affected -- either enhanced or 
suppressed -- by short-range interactions, and without knowing all 
parameters in the theory, we cannot calculate the exponents reliably.  
The oscillation at wavevector $(k_{F1}^{(0)} - k_{F1}^{(\pi)}, \pi)$ 
can be shown to be always suppressed beyond the mean field.
Finally, the zero momentum power law remains unchanged.

More details on the DBL[2,1]-phase can be found in Appendix~\ref{app:GT}.
In particular, we show that the two-boson Green's function exhibits 
some internal d-wave character, which originates from the non-trivial 
wavefunction signs.  It is these signs that prompted the name 
``D-wave'' Boson Liquid (DBL) in the 2D continuum setting,\cite{DBL} 
where the wavefunction goes through a sequence of signs $+-+-$ 
upon taking one particle around another.
However, such correlations do not necessarily mean that the system is 
near a d-wave-paired phase.  For example, we also examine a potential
instability of the DBL[2,1] driven by an allowed non-marginal
four-fermion term and find that the resulting phase is a boson-paired
liquid with an internal s-wave character.

Let us briefly mention the DBL[2,2] case, where in the mean field
both $d_1$ and $d_2$ have partial occupations of both the bonding and 
antibonding bands.  One can calculate various correlations in the 
mean field as before, and the structure is more rich since there is 
an additional band present.  
Some details are given in Appendix~\ref{subapp:DBL22instab}.
Beyond the mean field but ignoring non-marginal four-fermion terms,
this phase would have three gapless modes.  However, an analysis of 
allowed interactions suggests that this phase has a strong instability 
towards a boson-paired phase with an s-wave character.  
Remarkably, as we will see in the next section, a boson-paired phase 
with similar properties is found in the ring model in the regime of 
small interchain coupling $J_\perp$ where we initially hoped to find
the DBL[2,2] phase.
Thus, the slave-particle formulation and the gauge theory description 
solved via subsequent bosonization open a non-perturbative access 
to this interesting boson-paired phase.

\section{Frustrated $J$-$K$ Model on the 2-Leg Ladder}
\label{sec:JK}

\begin{figure}[t]
\centerline{\includegraphics[width=\columnwidth]{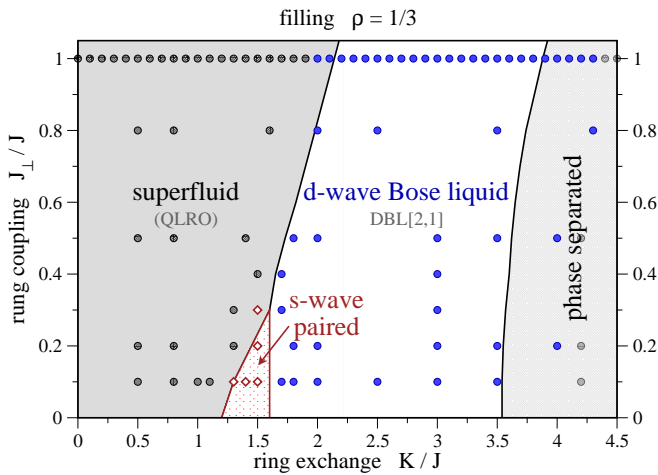}}
\caption{
(color online)
Phase diagram for the 2-leg boson system at filling $\rho=1/3$.
The symbols represent DMRG estimates with shaded grey circles
indicating the superfluid phase, the filled blue circles the 
d-wave correlated Bose liquid (DBL), and the open red diamonds the 
s-wave paired state.
The schematic line boundaries are obtained from VMC calculations.}
\label{fig:phased1}
\end{figure}


We study the ring model Eq.~(\ref{Hring}) on a 2-leg ladder with a 
boson hopping term $J$ that allows different amplitudes along and 
perpendicular to the chains:
\begin{eqnarray}
H_J = - J \sum_{\bf r} (b^\dagger_{\bf r} b^\pd_{{\bf r} + \hat{x}} + \Hc)
- J_\perp \sum_x (b^\dagger_{(x, 1)} b^\pd_{(x, 2)} + \Hc) ,
\end{eqnarray}
where ${\bf r}=(x,y)$ represent integer lattice sites.
We take $y=1, \dots, L_y$ with $L_y=2$ and $x=1, \dots, L_x$ with
$L_x$ the length of the 2-leg system.
The ring terms are associated with the square plaquettes of the 
2-leg ladder.

\begin{figure}[t]
\centerline{\includegraphics[width=\columnwidth]{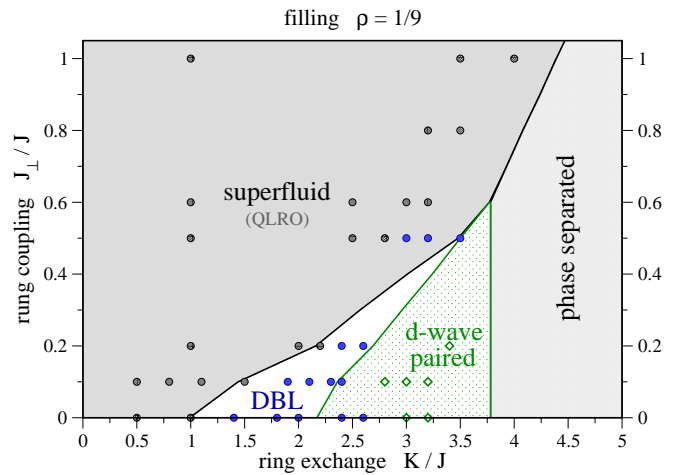}}
\caption{
(color online)
Phase diagram for the 2-leg boson system at filling $\rho=1/9$.
The symbols indicate DMRG estimates for the various phases
with shaded grey circles for the superfluid phase, 
filled blue circles for the DBL, and open green diamonds for the 
d-wave paired state.  
The schematic line boundaries are obtained from VMC calculations.
}
\label{fig:phased2}
\end{figure}

We use exact diagonalization (ED) and density matrix renormalization 
group (DMRG)\cite{dmrg, dmrg2} methods supplemented with trial 
wavefunction variational Monte Carlo (VMC)\cite{vmc} to determine the 
nature of the ground state of the Hamiltonian Eq.~(\ref{Hring}) at 
boson filling number $\rho = \frac{N_b}{L_x L_y}$, where $N_b$ is the 
total number of bosons in the system.
The obtained phase diagrams at densities $\rho=1/3$ and $1/9$ are shown 
in Fig.~\ref{fig:phased1} and Fig.~\ref{fig:phased2} respectively.  
These are typical results for the boson ring model 
away from half-filling.

Using ED and DMRG, we find four different quantum phases upon increasing
the ring-exchange coupling $K$ and varying the interchain coupling 
$J_\perp$.
At small $K$, the ``superfluid'' phase (SF) with ${\bf q} = {\bf 0}$
quasi-long-range order (QLRO) is stable for both filling numbers.  
The DBL phase, which is produced by the ring exchanges, dominates 
the intermediate parameter space at higher filling number $\rho=1/3$, 
while it occupies a substantially smaller region at $\rho=1/9$.  
Interestingly, at $\rho=1/3$ and small $J_\perp$, an s-wave paired phase 
emerges between the SF and DBL as a consequence of the competition 
between the boson hopping and ring-exchange terms.
At $\rho=1/9$ and small $J_\perp$, a d-wave paired phase is found 
adjacent to the DBL.  
Finally, for strong ring exchanges phase separation eventually wins near 
$K\sim 4J$ for both fillings, separating into a region with $\rho=1/2$ 
and an empty region $\rho=0$.
The characteristic features of each phase will be discussed in the 
following sections.

Our complementary VMC study is centered around the DBL phase.
The DBL wavefunction Eq.~(\ref{PsiDBL}) is constructed from the 
slave-particle formulation and thus allows us to compare between the 
exact ED/DMRG results and the gauge theory description.  
Jastrow SF and boson-paired trial wavefunctions 
(described in Appendix~\ref{app:PsiPaired}) are also considered 
in order to better understand the exact numerical results.
We note that this variational approach has been rather successful in 
one dimension, e.g., in the original studies of the 1D $t-J$ model.
\cite{HellbergMele, Chen, Yang}  We borrow some ideas from these
studies such as allowing the fermionic determinants to also play a 
Jastrow role via ${\rm det} \to {\rm sign(det)} |{\rm det}|^p$, 
where $p$ is a variational parameter.
In our ring model, the phase boundary estimated from VMC almost coincides
with the one determined from DMRG as indicated in Fig.~\ref{fig:phased1} 
for $\rho=1/3$ filling. 
At smaller filling number, DMRG and VMC also find the same topology of 
the phase diagram as given in Fig.~\ref{fig:phased2}.


We first describe measurements in the ground state.
Imposing periodic boundary conditions in the $\hat{x}$-direction in 
our numerical calculations, we can characterize the ED states by
a momentum $k_x$.  In our DMRG calculations we work with real-valued 
wavefunctions which gives no ambiguity when the system exhibits a 
unique ground state that caries zero momentum.  If the ground state 
carries nontrivial momentum $k_x$, its time-reversed partner carries 
$-k_x$, and the DMRG state is some real-valued combination of these.  
While the lattice-space measurements may depend slightly on the 
specific combination (but vanishingly in the large system limit), 
the momentum space measurements $n({\bf q})$ and $D({\bf q})$ described 
below are unique and insensitive to this.
In our DMRG calculations, more than 1500 states were kept in each 
block\cite{dmrg, dmrg2} to ensure accurate results, and the 
density matrix truncation error is of the order of $10^{-7}$.
The typical error for the ground-state energy is of the order of 
$10^{-5}$ for all systems we have studied.
The relative error in the correlation functions varies between
$10^{-4} - 10^{-2}$, depending on the type of the correlations
and the spatial distance between operators.

We measure boson correlator in the ground state
\begin{equation}
G_b({\bf r}, {\bf r}') = \la b_{\bf r}^\dagger b^\pd_{{\bf r}'} \ra
\end{equation}
and its Fourier transform
\begin{equation}
n({\bf q}) = \frac{1}{L_x L_y} \sum_{{\bf r}, {\bf r}'}
G_b({\bf r}, {\bf r}') e^{i {\bf q} \cdot ({\bf r} - {\bf r}')}
= \la b_{\bf q}^\dagger b^\pd_{\bf q} \ra ~,
\end{equation}
which is interpreted as mode ${\bf q}$ occupation number.

We also measure the boson density correlator
\begin{equation}
D_b({\bf r}, {\bf r}')  = 
\la (\rho_b({\bf r}) - \bar\rho)
    (\rho_b({\bf r}') - \bar\rho) \ra ~,
\end{equation}
where $\bar\rho$ is the average boson density.
The density structure factor is
\begin{equation}
D_b({\bf q}) = \frac{1}{L_x L_y} \sum_{{\bf r}, {\bf r}'}
D_b({\bf r}, {\bf r}') e^{i {\bf q} \cdot ({\bf r} - {\bf r}')} 
= \la \delta\rho_{-{\bf q}} \delta\rho_{\bf q} \ra ~.
\end{equation}
For both the boson or boson density correlations, a power law 
$1/x^p$ envelope in real space gives rise to a singularity 
$|\delta q|^{p-1}$ in momentum space.

Finally, we measure two-boson (pairing) correlator,
\begin{equation}
P_{2b}({\bf r}_1, {\bf r}_2; {\bf r}_1^\prime, {\bf r}_2^\prime) =
\la b^\dagger({\bf r}_1) b^\dagger({\bf r}_2) 
    b({\bf r}_1^\prime) b({\bf r}_2^\prime ) \ra ~,
\end{equation}
which creates a pair of bosons at ${\bf r}_1, {\bf r}_2$ and removes a 
pair at ${\bf r}_1^\prime, {\bf r}_2^\prime$.
This is useful for detecting boson-paired phases where the single 
boson correlator decays exponentially.
In the pairing correlator figures below, we present $P_{2b}$ for 
a fixed created $+45^\circ$ diagonal pair
$\{ {\bf r}_1=(0,1),\; {\bf r}_2=(1,2) \}$ near the origin,
while the removed pair is either a $+45^\circ$ diagonal
$\{ {\bf r}_1^\prime = (x,1),\; {\bf r}_2^\prime = (x+1,2) \}$
or a $-45^\circ$ diagonal pair
$\{ {\bf r}_1^\prime = (x,2),\; {\bf r}_2^\prime = (x+1,1) \}$.
We have measured the correlator for other pair orientations as well,
but find that the diagonal-diagonal correlations are the most 
distinguishing ones in the two paired phases.
When the amplitudes for the $\pm 45^\circ$ diagonals have the same
sign, we refer to this as s-wave character;
on the other hand, when the signs are opposite, we call this d-wave.
An extended discussion can be found in Appendix~\ref{app:GT} 
following Eq.~(\ref{Phipair2D}).

\subsection{Superfluid Phase}
\label{subsec:SF}

\begin{figure}
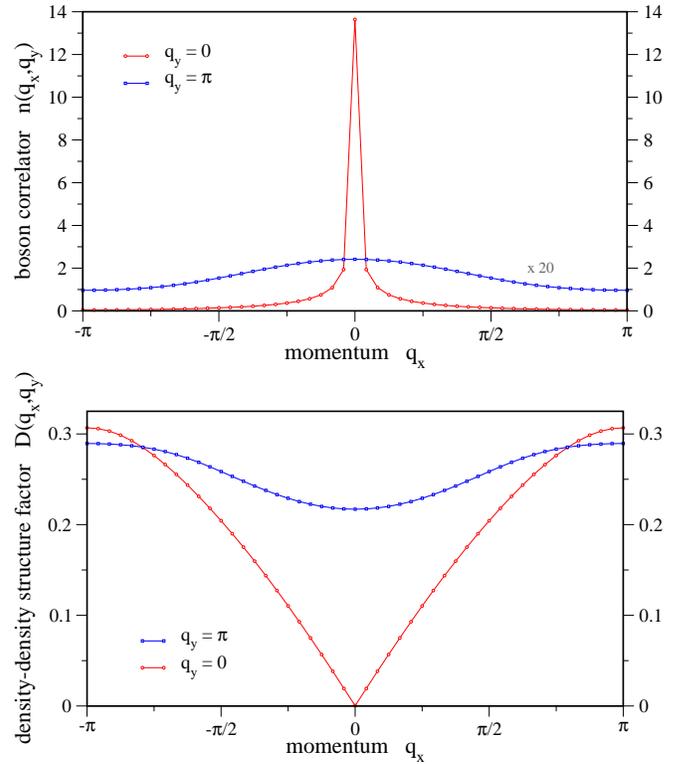

\centerline{\includegraphics[width=\columnwidth]{figures/SF_nq.eps}}
\centerline{\includegraphics[width=\columnwidth]{figures/SF_Dq.eps}}
\caption{
(a) The boson occupation number $n({\bf q})$ and (b) the density-density
structure factor $D({\bf q})$ for the 2-leg system at $\rho=1/3$,
$J_\perp=J$, $K=J$; the system length is $L_x=48$. 
The results are representative for the superfluid phase.}
\label{fig:sf_nq}
\end{figure}

\begin{figure}
\centerline{\includegraphics[width=\columnwidth]{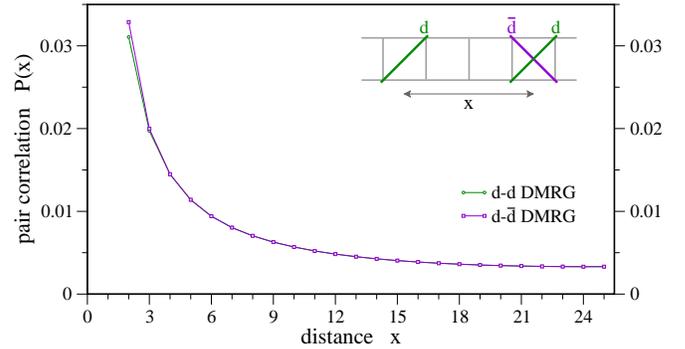}}
\caption{
The boson pairing correlations in real space for the SF system 
described in Fig.~\ref{fig:sf_nq}.
We show correlations for pairs on the diagonals $d$ and $\bar{d}$,
but roughly similar values are obtained for all pair orientations.
}
\label{fig:sf_pair}
\end{figure}

\begin{figure}
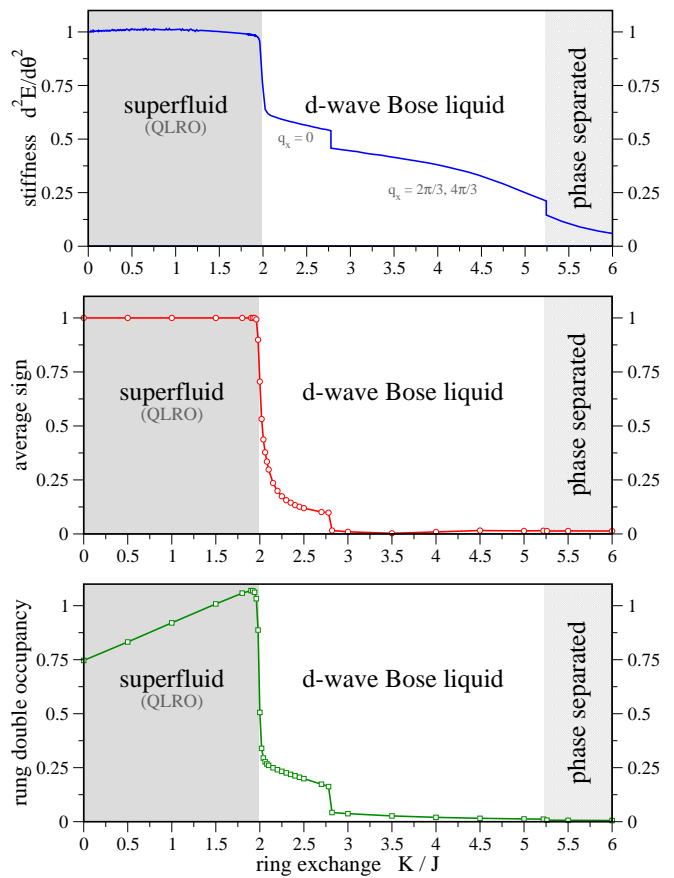

\centerline{\includegraphics[width=\columnwidth]{figures/ED_stiffness.eps}}
\vskip 2mm
\centerline{\includegraphics[width=\columnwidth]{figures/ED_sign.eps}}
\vskip 2mm
\centerline{\includegraphics[width=\columnwidth]{figures/ED_rung_occupancy.eps}}
\caption{
  Exact diagonalization results for a system of size $L_x=12$ at filling $\rho=1/3$ and $J_\perp = J$.
  Shown are the superfluid stiffness (top panel), the average sign of the ground-state
  wavefunction (medium panel) and the average double occupancy of rungs (lower panel)
  as a function of the ring exchange $K$.
}
\label{fig:ED_results}
\end{figure}

Let us first discuss the phase at very small $K$.  
The system is not frustrated in the $K=0$ limit and the ground-state
wavefunction (WF) components in the boson occupation number basis should 
all be of the same sign (positive) to gain the lowest hopping energy.
Interestingly, the ED calculations for small $L_x=12$ and $L_x=18$ 
systems find that this feature of the WF remains robust up to a finite 
value of $K \sim 2J$, close to the stability boundary of the SF phase.
This is illustrated in the medium panel of Fig.~\ref{fig:ED_results} 
where the average sign of coefficients in the ground-state wavefunction 
is plotted.
The boson occupation number $n(q_x, q_y)$ has a sharp peak at the 
wavevector ${\bf q} = (0, 0)$.  This is shown in Fig.~\ref{fig:sf_nq} (a)
for a system at $\rho=1/3$, $J_{\perp}=J$, $K=J$, and length $L_x=48$,
calculated using the DMRG method.  
For this system size, the peak value at ${\bf q} = (0,0)$ is about 
100 times larger than the value at wavevector ${\bf q} = (0,\pi)$.
This is consistent with the expected singular behavior 
$n(q_x, 0) \sim |q_x|^{p_b - 1}$ and non-singular $n(q_x, \pi)$ in the 
SF phase with 1D QLRO;
the boson correlator decays as $1/x^{p_b}$ in real space and we 
estimate $p_b \approx 0.25$.

As shown in Fig.~\ref{fig:sf_nq} (b) for the same system,
the boson density-density structure factor $D(q_x, 0)$ has a $|q_x|$ 
dependence around $q_x=0$ reflecting the global boson number 
conservation, while $D(q_x,\pi)$ is nonzero and has no visible 
singularities.

The pairing correlations in real space show power-law behavior, 
with nearly equal magnitudes for all pair orientations.
This is shown in Fig.~\ref{fig:sf_pair} where we create a 
$+45^\circ$ diagonal pair near origin and remove either a 
$+45^\circ$ diagonal pair or a $-45^\circ$ diagonal pair near $x$.
Furthermore, the exponent $p_{2b}$ of the power-law $1/x^{p_{2b}}$ 
in the pairing correlator is found to be $p_{2b} \approx 1$, 
which is about four times that for the single boson correlator,
$p_{2b} \approx 4 p_{b}$.  Such behavior of the pairing correlations is 
expected in the SF phase, which has QLRO in the single boson 
correlations.

These features characterize the SF phase.  It also has a finite 
$\rho_s L$, where $\rho_s$ is the ``superfluid stiffness'', 
which we can measure in our numerical calculations by imposing a 
small twist in the boundary conditions. 
Results from ED of a small system with $L_x=12$ are given in the
top panel of Fig.~\ref{fig:ED_results} and related DMRG measurements
for larger systems are in Fig.~\ref{fig:delta_E};
these are discussed in more detail in Sec.~\ref{subsubsec:rhos}.

\subsection{D-Wave Correlated Bose Liquid (DBL)} 
\label{subsec:DBL}

\begin{figure}[t]
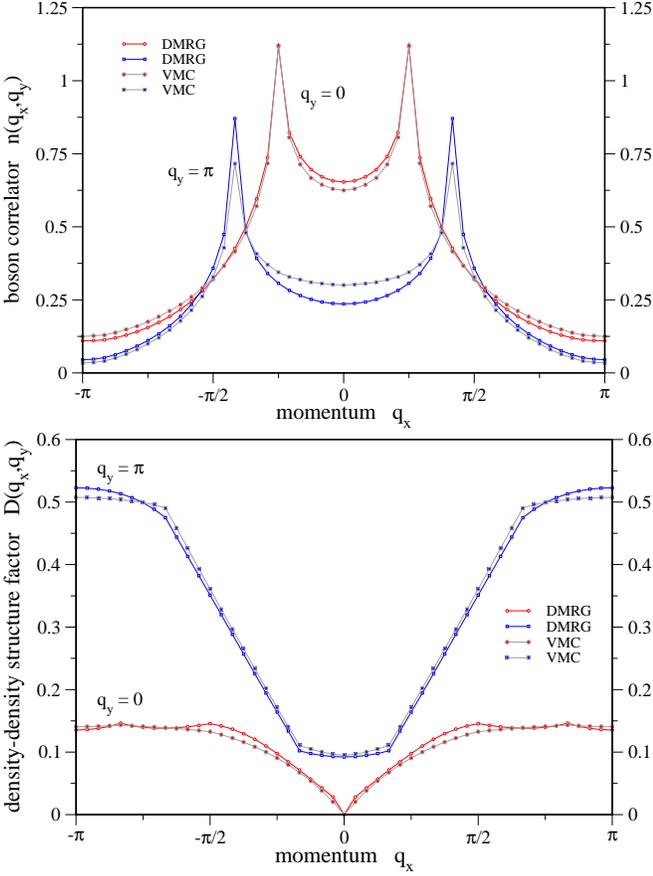

\centerline{\includegraphics[width=\columnwidth]{figures/DBL_nq.eps}}
\centerline{\includegraphics[width=\columnwidth]{figures/DBL_Dq.eps}}
\caption{
(a) The boson occupation number $n({\bf q})$ and (b) the density-density
structure factor $D({\bf q})$ for the 2-leg system at $\rho=1/3$,
$J_\perp=J$, $K = 3J$; the system length is $L_x=48$. 
The results are representative for the DBL phase.}
\label{fig:dbl_nq}
\end{figure}

As we further increase the strength of the ring-exchange $K$,
the system undergoes a (first order) phase transition where the 
stiffness $\rho_s$ drops steeply by a factor of two, see 
Figs.~\ref{fig:ED_results},\ref{fig:delta_E} and the subsection below.
There is also an abrupt change in the character of the wavefunction
sign which we can monitor in the ED analysis for small system sizes
by simple counting and shown in the middle panel in 
Fig.~\ref{fig:ED_results}.
\footnote{ 
In this small system with 8 bosons, Fig.~\ref{fig:ED_results},
there is also an abrupt change inside the DBL phase where the 
ground-state momentum jumps from $(k_x, k_y) = (0, 0)$ to 
$(k_x, k_y) = (2\pi/3, \pi)$ at around $K \approx 2.7~J$.  
We can understand these quantum numbers from specific orbital
occupations in the DBL wavefunction approach, and the change
of the total ground-state momentum arises from a change of 
orbital occupations in the $d_2$ bands.
Near $K \approx 2$ VMC finds the optimal DBL state to be
${\rm det}_1 [N_1^{(0)} = 8, N_1^{(\pi)} = 0]_{\rm abc} \times 
 {\rm det}_2 [N_2^{(0)} = 6, N_2^{(\pi)} = 2]_{\rm abc}$.
The label ``$\rm abc$'' stands for antiperiodic boundary conditions when
specifying the mean field orbitals; the total bosonic wavefunction 
satisfies periodic boundary conditions and carries momentum $(0, 0)$.
For the larger $K$, VMC finds the optimal DBL state 
${\rm det}_1 [N_1^{(0)} = 8, N_1^{(\pi)} = 0] \times
 {\rm det}_2 [N_2^{(0)} = 5, N_2^{(\pi)} = 3]$,
where now the orbitals are specified with periodic boundary conditions;
this state carries momentum $(k_x, k_y) = (2\pi/3, \pi)$.  
}
This new phase has unusual phenomenology and can be identified 
as the DBL[2,1] phase: There is no boson peak at ${\bf q} = (0, 0)$, 
but instead the boson occupation number $n(q_x, q_y)$ shows 
singular points at nonzero wavevectors $(\pm q_x^{(0)}, 0)$ and 
$(\pm q_x^{(\pi)}, \pi)$, as if there is a ``condensate'' of 
bonding and antibonding boson modes at these $q_x$.
Fig.~\ref{fig:dbl_nq} presents characteristic plots of the $n(q_x, q_y)$ 
and the density-density structure factor $D(q_x, q_y)$ deep in this 
phase for filling $\rho=1/3$ taken at $J_\perp = J$, $K = 3J$,
and a system length $L_x=48$.

\begin{figure}[t]
\centerline{\includegraphics[width=\columnwidth]{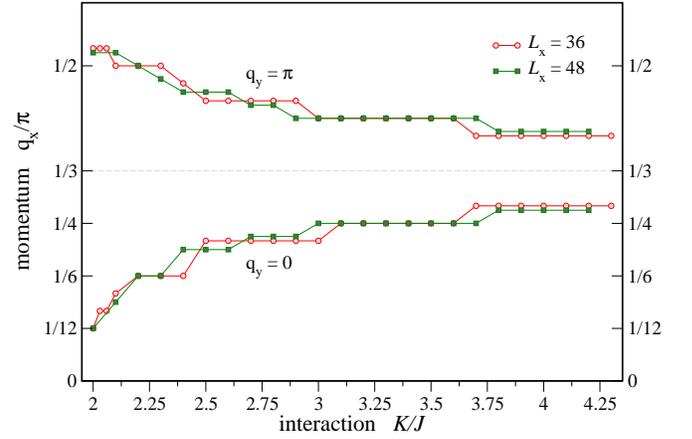}}
\caption{
Positions of the singular momenta in the DBL phase as a function of
the ring exchange $K$ for a system with $J_\perp = J$ and filling 
$\rho=1/3$.  
Data for systems with length $L_x = 36$ and $L_x = 48$ are given.
The momenta are determined from the peak values of $n(q_x, q_y)$ such 
as shown in the top panel of Fig.~\ref{fig:dbl_nq}.}
\label{fig:dbl_peak}
\end{figure}

The singular boson wavevectors $q_x^{(0)}$ and $q_x^{(\pi)}$ in general 
vary with parameters $K$ and $J_\perp$.  An example is shown in 
Fig.~\ref{fig:dbl_peak} for the system at $\rho=1/3$ with $J_\perp=J$ 
and varying $K$.
Just into the DBL phase near $K \gtrsim 2$, we find a small $q_x^{(0)}$ 
and a large $q_x^{(\pi)}$.  In the wavefunction construction, this
corresponds to rather different $k_{F1}^{(0)}$ and $k_{F1}^{(\pi)}$,
with significantly more $d_1$ fermions in the bonding than antibonding 
bands, see~Fig.~\ref{fig:bands}.
When we increase the ring coupling, $q_x^{(0)}$ decreases while 
$q_x^{(\pi)}$ increases, corresponding to moving the $d_1$ fermions 
from the bonding to antibonding levels and making the two bands 
more equally populated.   Interestingly, we find that the relation 
$q_x^{(0)} + q_x^{(\pi)} = 2 \pi \rho$ is satisfied, 
consistent with the prediction from gauge theory of a Luttinger-type  
theorem in the correlated boson phase.

The identification with the DBL is bolstered by a trial wavefunction
study.  The optimal such state for the system in Fig.~\ref{fig:dbl_nq} is
\begin{eqnarray}
\Psi_{\rm bos} &=& 
\left( {\rm det}_1 [N_1^{(0)} = 20, N_1^{(\pi)} = 12]_{\rm abc} 
\right)^{0.75} \\
&& \times
\left( {\rm det}_2 [N_2^{(0)} = 32, N_2^{(\pi)} = 0]_{\rm abc}
\right)^{0.1} ~.
\end{eqnarray}
The length of the system is $L_x=48$ and the total boson number is 
$N_b = 32$. 
The $d_1$ determinant is constructed from $N_1^{(0)} = 20$ bonding and 
$N_1^{(\pi)} = 12$ antibonding orbitals, while the $d_2$ determinant is 
constructed from bonding orbitals only.  To define the orbitals,
we use antiperiodic boundary conditions (``abc'') along the 
$\hat{x}$-direction for both $d_1$ and $d_2$, which gives closed shells 
for the above occupation numbers, while the physical boson wavefunction 
respects periodic boundary conditions and carries zero total momentum.  
The determinant powers are understood as
$({\rm det})^p \equiv {\rm sign(det)} |{\rm det}|^p$.
The $d_2$ determinant can be written explicitly as
\begin{equation}
{\rm det}_2[N_2^{(0)} = N_b, N_2^{(\pi)} = 0] \propto 
\prod_{i<j} \sin\frac{\pi (x_i - x_j)}{L_x} ~,
\label{detJW}
\end{equation}
and in particular prevents two bosons from sharing the same rung.
In fact, the exact ground states in the ED/DMRG have very small double 
occupancy of rungs deep in this phase, as illustrated in the 
lower panel of Fig.~\ref{fig:ED_results}.
The characteristic boson wavevectors in the above DBL state are
$q_x^{(0)} = k_{F2} - k_{F1}^{(0)} = 12\pi/48$ and
$q_x^{(\pi)} = k_{F2} - k_{F1}^{(\pi)} = 20\pi/48$.  These
bosons are created by occupying $d_1$ and $d_2$ orbitals on the 
opposite sides, which is motivated by the Amperean enhancement
of such composites as described earlier.
The overall match of the trial wavefunction results with the exact 
DMRG correlations is striking, reproducing the singular features
as illustrated in Fig.~\ref{fig:dbl_nq}.

Turning to the density structure factor $D(q_x, q_y)$, there are many 
features but they are also weaker than in the $n(q_x, q_y)$.  
For $q_y=0$, we expect a linear $|q_x|$-behavior near $q_x=0$ and 
signatures at $2 k_{F1}^{(0)} = 40\pi/48$, $2 k_{F1}^{(\pi)} = 24\pi/48$,
which we see, and also at $|2 k_{F2}| = 32\pi/48 \mod 2\pi$, 
which we do not see in either DMRG or VMC (we can understand this in the 
wavefunction because the power for the ${\rm det}_2$ is small and 
suppresses density correlations coming from this piece).
At $q_y=\pi$, we expect signatures at 
$k_{F1}^{(0)} + k_{F1}^{(\pi)} = 32\pi/48$ and at
$k_{F1}^{(0)} - k_{F1}^{(\pi)} = 8\pi/48$, both of which we see.
As discussed in Appendix~\ref{app:GT}, it is a complicated matter to 
predict the exponents of the different singularities, and given our
limited system size we do not attempt measuring these.
It is remarkable that the DBL wavefunction can reproduce the singular 
features, but we do not know at this point whether it can reproduce 
all exponents in general.

\begin{figure}[t]
\centerline{\includegraphics[width=\columnwidth]{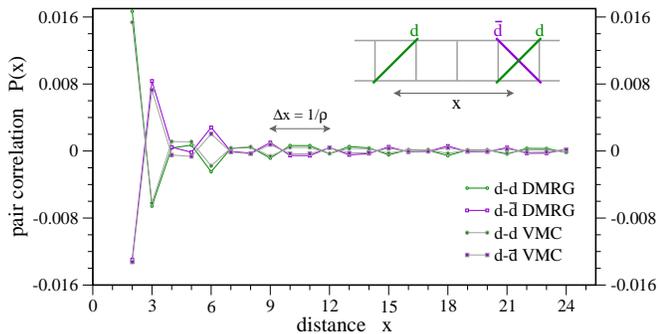}}
\caption{
The boson pairing correlations in real space for the DBL system 
described in Fig.~\ref{fig:dbl_nq} calculated by DMRG and VMC.
We show correlations for pairs on the diagonals $d$ and $\bar{d}$,
while values for other pair orientations are much smaller.
}
\label{fig:dbl_pair}
\end{figure}

Finally, in Fig.~\ref{fig:dbl_pair} we show the diagonal-diagonal
pairing correlator.  It is more short-ranged than in the SF phase 
as well as the two paired phases described below, but is characteristic 
of the DBL phase.
First, $P_{2b}(x)$ has robust sign oscillations at wavevector $2\pi\rho$.
Second, the $+45^\circ$ and $-45^\circ$ diagonal pairs have opposite 
signs; the nontrivial sign structure in the wavefunction gives some 
``D-wave''-correlated character to the boson liquid, which is 
observed in the two-boson correlator.

\subsubsection*{Detection of Boson Bonding and Antibonding Propagation\\ 
by Two Stiffnesses}
\label{subsubsec:rhos}

\begin{figure}[b]
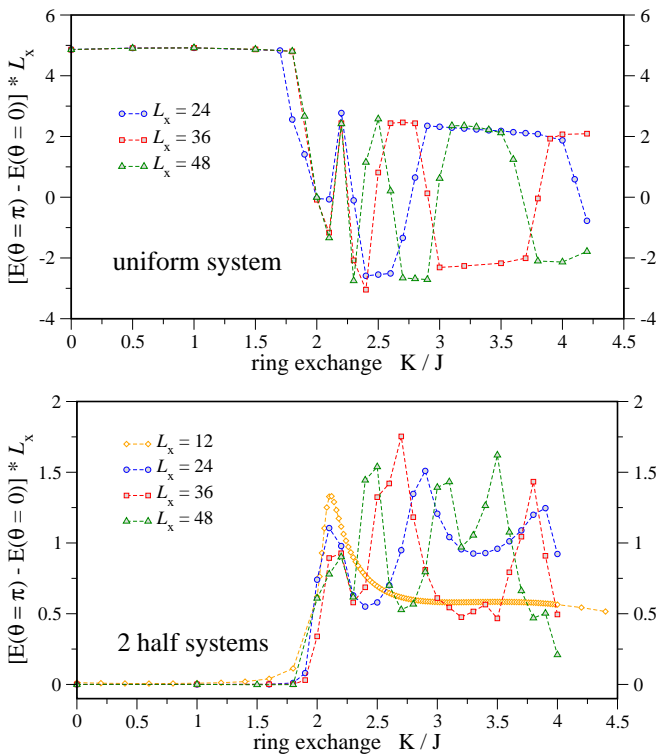

\centerline{\includegraphics[width=\columnwidth]{figures/StiffnessA.eps}}
\vskip 2mm
\centerline{\includegraphics[width=\columnwidth]{figures/StiffnessB.eps}}
\caption{
Detection of the boson propagation around the loop by applying a 
phase twist $\theta$ to the periodic boundary connection along the 
$\hat{x}$-direction.
The energy difference $\Delta E = E(\theta=\pi) - E(\theta=0)$,
multiplied by $L_x$,
(top panel) for a uniform system with $J_\perp = J$ for all bonds; 
(lower panel) for a specially designed system with $J_\perp = J$ in 
one half and $J_\perp = -J$ in the other half of the system.
}
\label{fig:delta_E}
\end{figure}

In this subsection we want to further highlight some characteristic 
difference between the SF and DBL phases.  
In the SF state, the wavefunction has QLRO at ${\bf q} = {\bf 0}$.
Imposing a twist boundary phase $\theta$ in the boson hopping term in 
the Hamiltonian (\ref{Hring}), one can measure the superfluid 
stiffness as $\rho_s = \partial^2 E / \partial \theta ^2$, which
is found to scale as $1/L_x$ in the SF phase.
This stiffness calculated from ED is shown for the small $L_x = 12$
system in the top panel of Fig.~\ref{fig:ED_results}.
For larger system lengths, the stiffness can alternatively be estimated 
by DMRG calculations of the energy difference for a $\theta=\pi$ twist, 
i.e., $\rho_s \sim \Delta E = E(\theta=\pi) - E(\theta=0)$.

In the top panel of Fig.~\ref{fig:delta_E}, we show $L_x \Delta E$ as a 
function of the ring exchange $K/J$ for several system sizes 
$L_x = 24,36,48$.  For isotropic hopping $J_\perp = J$, we find that
$L_x \Delta E$ is essentially constant in the SF phase, but drops 
abruptly at the SF to DBL transition. 
In the DBL phase, $L_x \Delta E$ shows rather irregular behavior
including sign changes when varying either $K$ or $L_x$.
For a fixed system length $L_x$, the more abrupt changes can be matched 
with the step-like changes of the singular momenta positions shown, e.g.,
in Fig.~\ref{fig:dbl_peak} for $L_x = 36,48$.
For fixed ring exchange $K$, the abrupt changes with varying $L_x$ are 
indicative of strong incommensurate correlations, where different $L_x$ 
offer varying degree of matching at the (twisted) periodic boundary:
$\Delta E > 0$ means that $\theta=0$ boundary conditions provide better
matching than $\theta=\pi$, while the situation is reversed if
$\Delta E < 0$.

A finite $|L_x \Delta E|$ in the large system limit in the above setting 
implies that there is a propagating ``bonding'' mode (i.e., with 
transverse momentum $q_y = 0$).
Based on our picture of the DBL phase, there must also be a propagating
``antibonding'' mode with $q_y = \pi$, which we would like to detect
directly and contrast with the SF phase where there is no such mode.

To this end, we design a system composed of two halves connected at
both ends, where the left half is a $(J, J_\perp = J, K)$-Hamiltonian
while the right half is a $(J, J_\perp = -J, K)$-Hamiltonian.
First, we note that a model with $J_\perp = -J$ can be related
to a model with $J_\perp = J$ by a transformation 
$b(x,y) \to (-1)^y b(x,y)$.  This ensures that the boson density is the 
same in the two parts.  Crucially, this transformation interchanges the 
notions of bonding and antibonding modes.

If the designed system is in the regime of the regular SF phase,
the gapless bonding bosons in the half with $J_\perp = J$ 
cannot penetrate into the other half where the gapless mode is 
anti-bonding, and vice versa.  Since the bosons are not able to 
propagate around the loop, a $\theta=\pi$ twist should not produce
significant change $\Delta E$ which will vanish exponentially with $L_x$.
On the other hand, in the DBL phase there are gapless modes of both 
bonding and antibonding type in either region, so that bosons can 
propagate around the loop, and the $\pi$-twist should change the energy 
by an amount proportional to $1/L_x$.
The ``bonding/antibonding propagation stiffness" $L_x \Delta E$ for the 
designed two-half system is shown in Fig.~\ref{fig:delta_E}(b) as a 
function of $K/J$.  
Indeed, we find a vanishingly small $L_x \Delta E$ in the SF phase 
and a rather irregular but nevertheless finite $L_x \Delta E$ in the 
DBL phase.
\footnote{
Note that $\Delta E$ in Fig.~\ref{fig:delta_E}(b) maintains a positive 
sign throughout the DBL phase, which can be understood roughly as 
follows:  A boson propagates via a bonding mode (with wavevector $q_x^{(0)}$) 
in one half and via an antibonding mode ($q_x^{(\pi)}$) in the other half.  
The total phase accumulated by going around the system is 
$(q_x^{(0)} + q_x^{(\pi)}) L_x/2 = \pi\rho L_x = 0 \mod 2\pi$
for $\rho=1/3$ and the shown $L_x$ multiples of 6.  Thus, the zero
twist boundary condition provides the best (crude) matching for all $K$.
When we tried different $L_x$ (not shown), the $\Delta E$ could be of 
either sign.
}

\subsection{Extended S-Wave Paired Boson Phase}
\label{subsec:s-paired}

\begin{figure}
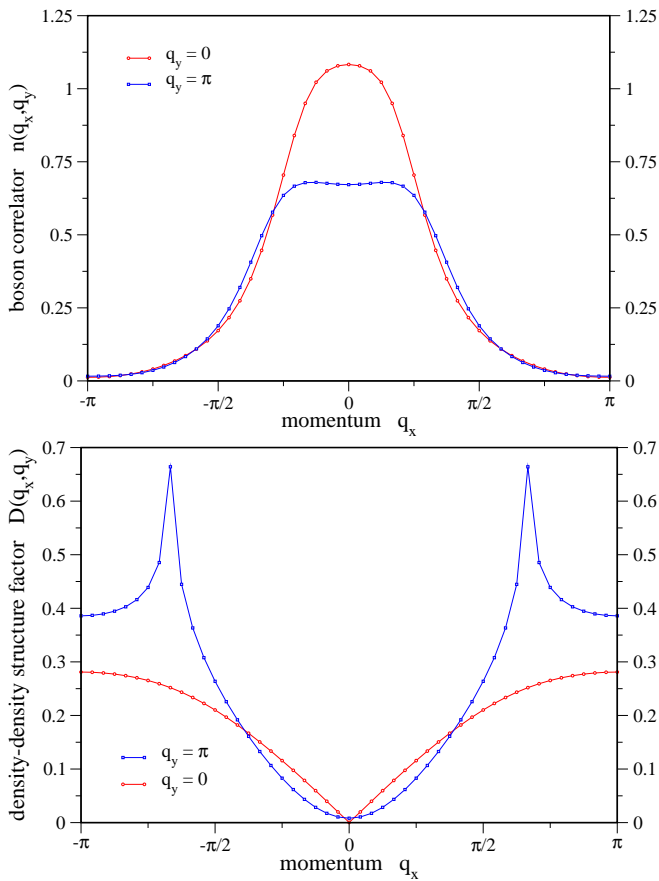

\centerline{\includegraphics[width=\columnwidth]{figures/SW_nq.eps}}
\centerline{\includegraphics[width=\columnwidth]{figures/SW_Dq.eps}}
\caption{
(a) The boson occupation number $n({\bf q})$ and (b) the density-density
structure factor $D({\bf q})$ for a 2-leg system at $\rho=1/3$,
$J_\perp=0.1J$, $K=1.4J$; the system length is $L_x=48$. 
The results are representative for the s-wave paired phase.}
\label{fig:sw_nq}
\end{figure}

For our 2-leg ladder system one might expect to find the DBL[2,2] phase 
if the interchain coupling $J_\perp$ is much smaller than the 
coupling $J$ along the chains: 
When the boson hopping between the chains is reduced, one expects that
the fermion hopping between the chains is also reduced for both species,
which in turn could bring the antibonding band of the $d_2$ fermions to 
cross the chemical potential as shown in the bottom right panel of 
Fig.~\ref{fig:bands}.
DMRG studies of the ring model at filling $\rho=1/3$ and small 
$J_\perp \leq 0.3$ indeed find a phase that is distinct from both the 
DBL[2,1] and SF as shown in the phase diagram Fig.~\ref{fig:phased1} 
around $K\sim 1.4J$.  However, our measurements indicate that this 
phase is not a DBL[2,2], but a paired phase.  
As shown in Fig.~\ref{fig:sw_nq} for a system at $\rho=1/3$ with 
$J_\perp=0.1J$, $K=1.4J$, and length $L_x=48$, this phase is 
characterized by a quite flat region in the boson occupation number 
$n(q_x,q_y)$ near $q_x=0$ without any singular points, which is 
different from both the SF and DBL phases. 
On the other hand, the density-density structure factor $D(q_x, q_y)$
has a singular behavior at a wavevector $(2\pi\rho, \pi)$.
This singular behavior can be approximated by 
$\sim |\delta q_x|^{p_\rho - 1}$, with $p_\rho$ between $1.33 - 1.50$ 
providing the best fit to the data, which corresponds to power law 
envelope $x^{-p_\rho}$ in real space.
[Note that when $J_\perp=0$ the number of bosons in each chain is 
independently conserved and $D(q_x, \pi)$ must vanish at $q_x=0$, 
which explains the small value of $D(q_x=0, \pi)$ extracted by DMRG 
when $J_\perp = 0.1J$ in Fig.~\ref{fig:sw_nq}.]

\begin{figure}[b]
\centerline{\includegraphics[width=\columnwidth]{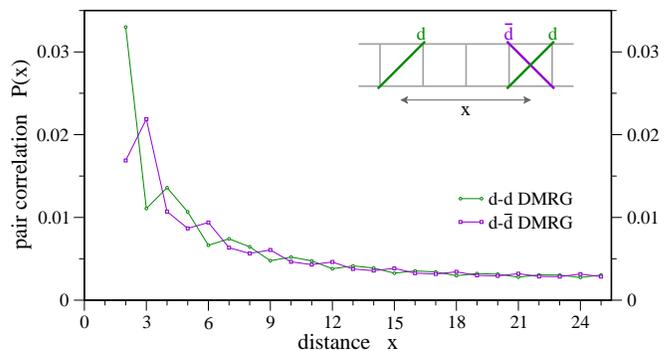}}
\caption{
The boson pairing correlations in real space for the same s-wave
paired system described in Fig.~\ref{fig:sw_nq}.  
We show diagonal pairs but roughly similar correlations are obtained 
also for pairs on the rungs, while correlations for pairs in the 
chains are significantly smaller.}
\label{fig:sw_pair}
\end{figure}

The observed phase can be identified as an s-wave boson paired phase 
with QLRO pairing correlation as shown in Fig.~\ref{fig:sw_pair}.  
The pairing correlation $P_{2b}(x)$ in this phase is strongly 
enhanced at large distances $x$ when compared to the correlation in 
the DBL phase; the values are roughly similar for pairs straddling the 
chains but smaller for pairs inside the chains.
The decay of $P_{2b}(x)$ can be fitted by a power law behavior 
$\sim 1/|x|^{p_{2b}} + 1/|L_x-x|^{p_{2b}}$ with $p_{2b} \approx 1$.

We have also checked the single boson correlation in the real space:
It is much reduced compared to the nearby SF and DBL phases and decays 
very quickly -- essentially exponentially; in particular, it is 
much smaller than the boson-pair correlation beyond few lattice spacings.
We then conclude that there is a gap to single-boson excitations, 
but not to pairs.
This identification of the phase is also supported by a VMC study using 
paired boson wavefunctions described in Appendix~\ref{subapp:PsiSW}, 
which improve the energetics over that of the trial DBL states and 
reproduce qualitatively the DMRG correlations.

Interestingly, the s-wave paired phase can be regarded as an instability 
of the DBL[2,2] phase.  In Appendix~\ref{subapp:DBL22instab}, we argue 
that without specially adjusted short-range interactions, the DBL[2,2] is
unstable once the gauge fluctuations are included, and the most natural 
outcome is a paired phase with the same phenomenology as described above.
In particular, the pairs are s-wave, carry zero momentum, and show QLRO 
with power law $x^{-p_{2b}}$, while we also predict power law charge 
correlations at wavevector $(2\pi\rho, \pi)$ with $x^{-p_\rho}$.
In Appendix~\ref{subapp:DBL22instab} we also predict that the two
exponents are inverses of each other, $p_{2b} p_{\rho} = 1$.
The DMRG estimates of the power laws differ somewhat from this, 
but they are hard to make accurate given the slow decay of the 
correlations and our limited system sizes.

\subsection{D-Wave Boson Paired  Phase}
\label{subsec:d-paired}

\begin{figure}
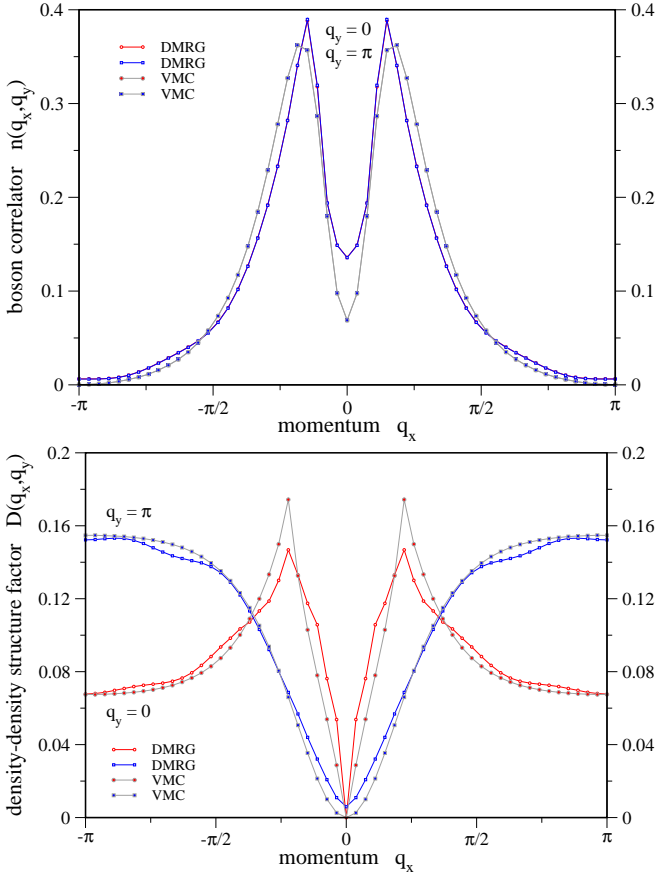

\centerline{\includegraphics[width=\columnwidth]{figures/DW_nq.eps}}
\centerline{\includegraphics[width=\columnwidth]{figures/DW_Dq.eps}}
\caption{
(a) The boson occupation number $n({\bf q})$ and (b) the density-density
structure factor $D({\bf q})$ for a 2-leg system at $\rho=1/9$,
$J_\perp = 0.1J$, $K=3J$; the system length is $L_x=54$. 
The results are representative for the d-wave paired phase.}
\label{fig:dw_nq}
\end{figure}

\begin{figure}
\centerline{\includegraphics[width=\columnwidth]{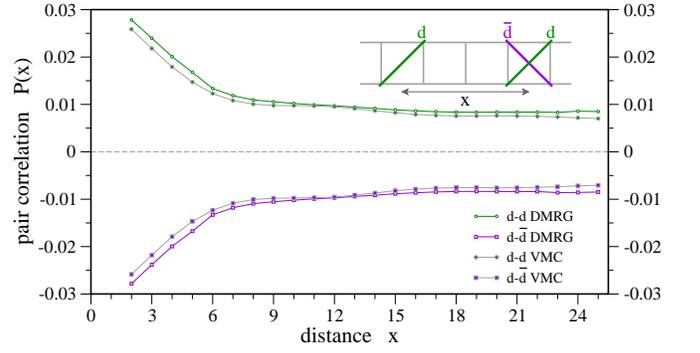}}
\caption{
The boson pairing correlations in real space
for the same system described in Fig.~\ref{fig:dw_nq}.}
\label{fig:dw_pair}
\end{figure}

Consider now bosons at small filling $\rho=1/9$, Fig.~\ref{fig:phased2}.
As we further increase $K$, the DBL phase in the intermediate range 
becomes unstable and a new phase emerges.
As shown in Fig.~\ref{fig:dw_nq} for a system at filling $\rho=1/9$ with 
$J_{\perp}=0.1J$, $K=3J$, and length $L_x=54$,  
this phase is characterized by $n(q_x,0) \approx n(q_x,\pi)$,
which is satisfied very accurately and is indicative of 
strongly suppressed boson correlator between the chains.  
We think that the peaks in $n({\bf q})$ are non-singular and represent 
short-range boson correlations, while $G_b({\bf r})$ decays exponentially
on long distances: 
While we cannot ascertain the exponential decay with our system sizes, 
the single boson correlator is already more than an order of magnitude 
smaller and decays much faster than the pair correlator described below.
This phase is further identified as a d-wave boson paired phase with 
power-law pairing correlation as shown in Fig.~\ref{fig:dw_pair}. 
The diagonal pairing correlations $P_{2b}(x)$ show opposite signs 
depending on the relative orientation of the two pairs: 
positive when both are $+45^\circ$ diagonals and negative when one is 
$+45^\circ$ and the other $-45^\circ$, which is consistent with the 
d-wave pairing picture.
We find power law correlation $P_{2b}(x) \sim 1/x^{p_{2b}}$ with 
exponent $p_{2b} \approx 0.85$ 

The identification of this as d-wave paired phase is supported by the 
reasoning at low densities described in Appendix~\ref{subapp:BoundState}.
Thus, for a pair of bosons on an otherwise empty ladder, there is a 
transition at $K/J = 2$ where it becomes favorable for bosons to form a 
molecule with internal d-wave character.
If we now have a system of bosons at small density, it is natural that 
there will be a parameter range where the bosons are paired and the 
resulting molecules form a Luttinger liquid.

To bolster this identification, we also studied such states 
variationally.
As described in Appendix~\ref{subapp:PsiSW}, we construct a trial 
wavefunction as a product of a Pfaffian and a determinant,
since these are simple to work with in VMC.
Optimal trial state for the system in Fig.~\ref{fig:dw_nq} is
\begin{eqnarray}
\Psi_{\rm bos} &=& {\rm Pf}_1 \times
\left( {\rm det}_2 [N_2^{(0)} = 12, N_2^{(\pi)} = 0] \right)^{0.3} ~.
\end{eqnarray}
The system length is $L_x = 54$ and the boson number is $N_b = 12$.
The Pfaffian is for an antisymmetric (spinless fermion) pair-function 
$g(x,y; x',y') = \delta_{y\neq y'} {\rm sign}(y-y') e^{-|x-x'|/\xi}$
that connects only sites on different chains; the optimal 
``size of the pair'' is $\xi=1.8$.  This is a ``$p_y$'' pairing
and when composed with a ``$p_x$'' character present in the determinant
this gives a d-wave character to the boson pairs.
Strictly speaking, the above state is valid only in the $J_\perp=0$ 
limit, since it prevents fluctuation of the boson number in each chain. 
Still, as we can see from Fig.~\ref{fig:dw_nq}, the variational 
wavefunction reproduces the DMRG correlations surprisingly well in 
this phase.  The boson momentum distribution function for the 
trial state is (by construction) non-singular, so that the close 
agreement with the DMRG found for $L_x = 54$ strengthens the 
identification of this phase as a d-wave paired.

Finally we mention the signatures in the boson density shown in
the lower panel in Fig.~\ref{fig:dw_nq}.
$D(q_x, 0)$ shows the familiar $|q_x|$ at small momenta and also a 
singularity at $2\pi\rho$.  The latter gives a wavelength equal to the
spacing between the pairs, which is expected in a Luttinger liquid of 
such molecules.  The density singularity is not too strong and is 
in line with the slow decay of the pair correlator.

\subsection{Phase Separation}
As we further increase the ring exchange towards $K \gtrsim 4J$, the 
uniform phase becomes unstable to phase separation.  For such large $K$, 
the boson density will separate into half-filled and empty regions, 
which is energetically favored by the ring exchange term.
In the ED of the small $L_x=12$ system, phase separation is identified 
to happen at $K \gtrsim 5J$ where all the lowest energy states from 
different momentum sectors become nearly degenerate.
In the DMRG calculation, the phase separation is indicated by a 
nonuniform boson density in the obtained ground state, which breaks 
the translational symmetry due to small perturbations from the cut-off 
of the Hilbert space in DMRG.
The phase separation occurs around $K \gtrsim 4J$ as observed from DMRG.
Finally, the phase separation can also be studied in VMC by measuring
trial energies for different boson densities and performing Maxwell 
construction (this is described in more detail, e.g., in 
Ref.~\onlinecite{DBL} Sec.~VII for the 2D ring model).

The tendency towards phase separation for large $K$ is expected since 
the ring term provides effective attraction between bosons, see 
Appendix~\ref{subapp:BoundState} and Ref.~\onlinecite{DBL} Sec.~VII.
Also, in the large $K$ limit, we can solve the ladder Hamiltonian 
exactly and find phase separation -- this is shown in 
Appendix~\ref{subapp:Konly}.
Nevertheless, for intermediate $K$ the various phases discussed above 
are stabilized by the boson hopping.

\section{Conclusions and Future Directions}
\label{sec:concl}

An obvious virtue of searching for the fingerprint of putative 
2D gapless quantum phases in quasi-1D models is the relative numerical 
and analytic tractability of the latter.
An important requirement of this approach is that the 2D quantum phase 
has singular correlations along surfaces in momentum space.
In this case the number of 1D gapless modes present on an $N$-leg ladder 
will necessarily scale linearly with $N$.
While the particular fingerprint of the 2D quantum phase can already be 
present on the 2-leg ladder, as was the case for the DBL phase 
studied exhaustively in this paper, a compelling approach to 2D will 
certainly require analyzing models with larger $N$.

Already when $N=3$ exact diagonalization studies become problematic 
since finding the ground states of boson or spin models with more than 
$36$ sites becomes virtually impossible.  Thus a $3$-leg ladder of 
length $L=12$ is already near  the limit.
DMRG is much more promising, and past studies on spin models 
reveal that convergence to the ground state is possible for lengths 
$L$ of order of a $50-24$  with  $N=3-6$, respectively.

Variational Monte Carlo for Gutzwiller-type wavefunctions 
constructed as products of determinants or Pfaffians as considered 
in this paper is possible for significantly larger $N$, and equal time 
correlation functions can be computed with a fair degree of accuracy.  
But VMC is only as good as the variational wavefunction being employed, 
and the quality of the wavefunction is hard to assess in the absence of 
other more accurate methods.  
The ground-state energy of all physically reasonable Hamiltonians is 
sensitive almost entirely to very short-range correlations,
whereas for gapless quantum systems it is usually the longer range
correlations which are necessary in order to distinguish
between various competing phases.  Quantum states in 2D with gapless
critical surfaces in momentum space are somewhat of an exception
since the location of such surfaces is presumably due to short-range 
correlations in the signs and amplitudes of the wavefunction.
But the precise singular structure on such surfaces reflects longer 
range correlations in the wavefunction.

Gauge theory\cite{LeeNagaosaWen} seems to offer one of the few 
analytic approaches to access putative gapless spin-liquid type 
quantum phases in 2D that have no quasi-particle description.
And for mean field states having a spinon Fermi sea, 
a controlled inclusion of the gauge fluctuations is problematic, 
even in the simplest case of a compact $U(1)$ gauge field.
Accounting for the instanton (monopole) events which reflect
the compact nature of the gauge field is exceedingly difficult.
Even for a non-compact $U(1)$ gauge field coupled to a Fermi sea,
while there does exist an RPA-type approach,
\cite{LeeNagaosa, Polchinski, Altshuler, LeeNagaosaWen}
it involves uncontrolled and worrisome approximations.

But for the ladder systems, gapless fermions (spinons) coupled to
a $U(1)$ gauge field is eminently tractable, although this fact has been 
exploited rather infrequently.\cite{KimLee, Hosotani, Mudry, Ivanov}
In 1D the gauge field has no transverse component and the bosonization
method can be employed, leading to a gaussian effective field
theory for the low energy excitations.
This is another distinctive advantage of studying ladder systems
relative to their 2D counterparts.

\subsection{DBL Phase on the N-Leg Ladder}

One extension of this paper involves studying the boson ring Hamiltonian 
on ladders with $N=3,4,5$ and larger.  Already for $N=3$ there are a 
number of new features that are not present on the 2-leg ladder.  
First, one can consider either periodic or anti-periodic boundary 
conditions for the bosons in the rung-direction, and it is quite 
plausible that the phase diagram and the presence/absence of the 
DBL phase will depend on this choice.
Second, while our search for a DBL phase with three gapless modes 
on the 2-leg ladder was unsuccessful due to instabilities of the DBL[2,2]
variational state, it seems likely that for $N \geq 3$ a DBL phase with 
more gapless modes than $N$ should be accessible.  This is interesting 
since in such a case at least two of the 1D modes would have the same 
transverse momentum.

For a general $N$-leg ladder one can construct variational states of the 
form DBL[$n$, $m$] with integer $n,m$ satisfying $N \geq n \geq m > 0$ 
(with $n, m$ the number of partially filled bands for the $d_1, d_2$ 
partons, respectively).  The number of 1D gapless modes in such a phase 
is $n+m-1$, so that DBL[3,2] for $N=3$, say, would have four 1D modes.
If $m=1$ the $d_2$ fermion has only one band partially occupied, and
assuming the band has $k_y=0$, the corresponding variational wavefunction
vanishes whenever two or more bosons occupy a rung.  In this case of 
no-double rung occupancy, the determinant for the $d_2$ fermion can be 
viewed as a (strictly) 1D Jordan-Wigner string multiplying the $d_1$ 
determinant.  Whenever $m > 1$ this is no longer the case, and the $d_2$ 
determinant would have a more subtle effect on the signs of the boson 
wavefunction.  It would thus be desirable to access a DBL phase with 
$m > 1$.
As detailed in Appendix~\ref{subapp:DBL22instab}, the gauge theory 
suggested an instability of the DBL[2,2] state for $N=2$ due to a 
rather special ``nesting" condition present for $n=m=2$.  
For $n > m \geq 2$ this will not generally be the case so we expect that 
such DBL phases should be more easily accessible for $N=3,4,\dots$.
It would also be interesting to use the DMRG to access some dynamical
information about the DBL, which should be possible at least for $N=2,3$
For $N \to \infty$ the ladder model recovers the full symmetry of the 
2D square lattice, and in future work it would be desirable to study
how to approach this limit.

\subsection{SU(2) Invariant Spin Models with Ring Exchange}

It would be most interesting to search for possible 2D spin liquid
phases with singular surfaces in momentum space in models possessing 
SU(2) spin symmetry. Recently, several authors
\cite{ringxch, SSLee, Senthil} have suggested that the putative 
spin liquid observed in $\kappa$-(ET)$_2$Cu$_2$(CN)$_3$ is of such 
character, and have proposed that a mean field state with a 
Fermi surface of spinons is an appropriate starting point.  
Variational energetics on the corresponding Gutzwiller projected 
spinon Fermi sea have been performed on the triangular lattice 
Heisenberg antiferromagnet with a four-site cyclic ring exchange term, 
a model argued to be relevant for a Mott insulator such as 
$\kappa$-(ET)$_2$Cu$_2$(CN)$_3$ that has a comparatively small 
charge gap.  

It should be possible to study the Heisenberg plus ring exchange
spin Hamiltonian on the triangular strip to search for signatures of the 
proposed parent 2D spin liquid.  In the absence of ring exchange this 
model is equivalent to the 1D Heisenberg model with both first and 
second neighbor interactions, $J_1, J_2$, and has been explored using 
DMRG in earlier work.\cite{zigzag, Nersesyan}
The phase diagram as a function of $J_2/J_1$ appears to have two phases:
a Bethe-chain phase stable for small $J_2/J_1$ with one gapless mode 
and a fully gapped dimerized state for intermediate and large $J_2/J_1$;
the gap decreases exponentially for large $J_2/J_1$ where we have
nearly decoupled legs ($J_2$) of the triangular ladder 
(with interchain $J_1$).

With a 4-site ring exchange term (coupling $K$) the triangular strip 
model is characterized by two dimensionless parameters, 
$J_2/J_1, K/J_1$.  
Are there any additional phases in this phase diagram?
The spin liquid phase on the triangular strip that descends from the 
(putative) 2D spinon Fermi surface state should have 3 gapless modes
(2 for spin, 2 for the two legs, minus one from the gauge constraint)
and corresponding incommensurate spin correlations.
Variational Monte Carlo together with a gauge theory analysis can 
provide a detailed characterization of this state, which could then be 
compared with DMRG, precisely as was done in this paper for the 
boson ring model.  
If present, one expects the spin liquid phase to appear
for intermediate values of both $J_2/J_1$ and $K/J_1$.

In addition, one might study a half-filled Hubbard model on the 
triangular strip, which should exhibit a metal insulator transition at 
intermediate coupling, $U/t$.  Of particular interest is the nature
of the quantum state just on the insulating side of the Mott transition.
If the Mott transition is weakly first order, there will be substantial 
charge fluctuations and ring exchange interactions in this part of the 
phase diagram, and possibly a spin liquid phase,
in addition to phases that have already been identified
(see Refs.~\onlinecite{Daul, Capello, Japaridze} and references
therein).

\subsection{Itinerant Electrons}

What are the prospects of using ladders to approach non-Fermi liquid 
phases of 2D itinerant electrons?
One complication is as follows.  Imagine a weak coupling Hubbard model
on the square lattice (at densities well away from special commensurate 
values) which has a conventional 2D Fermi liquid ground state.
On an $N$-leg ladder the free Fermi points present for $U=0$ would be 
converted into Luttinger liquids.  The jump discontinuity in the 
momentum distribution function at each Fermi point would be lost, 
but singularities would remain, characterized by some Luttinger liquid 
exponents.  Moreover, the location of the singularities would still 
satisfy the Luttinger sum rule.
Now imagine some strong coupling 2D electron Hamiltonian with a 
non-Fermi liquid phase that has a residual Luttinger Fermi surface
but with $Z=0$, analogous to a 1D Luttinger liquid.
The corresponding $N$-leg ladder descendant would be qualitatively 
indistinguishable from the phase of the weak coupling Hamiltonian.
Conversely, the presence of a Luttinger-satisfying Fermi surface on an 
$N$-leg ladder would not enable one to distinguish between the 
two different 2D phases, one a Fermi liquid and the other not.

On the other hand, the $N$-leg ladder descendant of a 2D non-Fermi 
liquid ground state with momentum space singularities that violate the 
Luttinger sum rule (surfaces with the ``wrong" volume, or perhaps even 
arcs) would have qualitatively distinct signatures.
Several recent papers\cite{DBL, Senthil} have proposed such 2D 
non-Fermi liquid phases,
and it would be extremely interesting to find evidence for their
ladder descendants.  The wavefunction for one of these phases 
was constructed\cite{DBL} by taking the product of a free fermion 
determinant and the DBL wavefunction in Eq.~(\ref{PsiDBL}). 
This phase, which inherits the d-wave sign structure from the bosonic DBL
wavefunction, was called a d-wave metal and would exhibit distinctive 
signatures if present on a ladder system.
A possible Hamiltonian that might possess the d-wave metal can be 
expressed by adding an ``itinerant-electron-ring" term to the usual 
square lattice $t-J$ model:
$H = H_{t-J} + H_{\rm ring}^{\rm el}$ with,
\begin{equation}
H_{\rm ring}^{\rm el} = K_{\rm el} \sum_\square 
({\cal S}_{1,3}^\dagger {\cal S}^\pd_{2,4} + \Hc) ,
\end{equation}
where the sites $1,2,3,4$ are taken to run clockwise around an 
elementary square plaquettes and the summation runs over all packets.
Here we have defined an electron singlet creation operator:
\begin{equation}
{\cal S}_{i,j}  = (c_{i\up}^\dagger c_{j\dn}^\dagger 
- c_{i\dn}^\dagger c_{j\up}^\dagger) / \sqrt{2} .
\end{equation}
The ring term rotates a singlet on the $2-4$ diagonal into
a singlet on the $1-3$ diagonal.
It would be interesting to study this or other such strong coupling 
Hamiltonians on the $N$-leg ladder.

\subsection{Conclusions}
To conclude, in this paper we have initiated a study of $N$-leg 
ladder systems that are descendants of candidate 2D quantum phases 
with low-energy excitations residing on singular surfaces.  
Motivated by one such proposal for the so-called DBL phases of 
un-condensed itinerant bosons,\cite{DBL} we have studied the 2-leg model 
with frustrating ring exchanges using exact numerical approaches.
We have indeed found the DBL[2,1] ladder version arising prominently in 
the phase diagram.
We have also searched for the DBL[2,2] version but from the gauge
theory analysis concluded that it is likely unstable due to special 
kinematics on the 2-leg ladder and instead gives rise to a boson-paired 
phase with extended s-wave character in the pairs.
This paired phase is in fact realized in our model, while the
DBL theory gives us tools to understand its properties.
While our focus has been on the DBL ideas, we have explored the full
phase diagram of the specific ring model in fair detail and found also 
other strong-coupling phases such as the above s-wave paired phase and 
the d-wave paired phase.  The latter is not directly accessible from
the DBL theory but is characteristic of the binding tendencies in our
ring terms, which eventually cause the bosons to phase-separate for
large ring exchanges.
We hope to pursue similar ideas in the $N \geq 3$ ladders and also
in SU(2) spin and itinerant electronic models with ring exchanges, 
which are particularly exciting.

\acknowledgments
We would like to thank L. Balents, T.~Senthil, A.~Vishwanath 
for useful discussions.
This work was supported by
DOE grant DE-FG02-06ER46305 (DNS), 
the National Science Foundation through grants 
DMR-0605696 (DNS) and DMR-0529399 (MPAF), 
and the A.~P.~Sloan Foundation (OIM).
Some of our numerical simulations were based on the ALPS libraries \cite{ALPS}.


\appendix

\section{Gauge Theory Description and Solution by Bosonization}
\label{app:GT}

A faithful formulation of the physical system in the slave particle
approach Eq.~(\ref{d1d2}) is a compact U(1) lattice gauge theory.  
We set this up on the 2-leg ladder as follows.  
Denote the vector potential components on the links of the two chains as 
$a_{I,x}$ and $a_{II,x}$, and on the rungs as $a_y$.
In the Euclidean path integral, denote the temporal components associated
with the sites on the two chains as $a_{I,\tau}$ and $a_{II,\tau}$.
The action for the gauge field contains 
$-\cos(a_{I,x} - a_{II,x} + \nabla_x a_y)$, 
$-\cos(a_{I,\tau} - a_{II,\tau} + \nabla_\tau a_y)$,
and $-\sum_{l=I,II} \cos(\nabla_\tau a_{l,x} - \nabla_x a_{l,\tau})$.
For each $l=I,II$, the last term gives a 1+1D gauge action,
and if the cosine is interpreted as a Villain cosine, we can replace 
it with $(\nabla_\tau a_{l,x} - \nabla_x a_{l,\tau})^2$ and treat 
$a_l$ as non-compact with no change in the results.
Choosing the gauge $a_y=0$, the combination $a_I - a_{II}$ is massive, 
while the combination $a_I + a_{II}$ is massless.

Having described the gauge sector, we now consider the fermions $d_1$ 
and $d_2$ with the mean field bands as in Fig.~\ref{fig:bands} and 
take the continuum limit using the bonding $d_\alpha^{(0)}$ and 
antibonding $d_\alpha^{(\pi)}$ fields near the corresponding 
Fermi points $P = R/L = +/-$.  
These fields couple to $a = (a_I + a_{II})/2$ as to the usual 
gauge field, while the massive combination $a_I - a_{II}$ can be 
integrated out.  The continuum Hamiltonian density is
\begin{eqnarray}
h &=& h_{kinetic} + h_{4ferm} ~, \\
h_{kinetic} \!\!&=&\!\!\!
\sum_{\alpha, k_y, P} P v_{F\alpha}^{(k_y)}
d_{\alpha P}^{(k_y)\dagger} (-i\partial_x - e_\alpha a_x) 
d_{\alpha P}^{(k_y)} ,
\end{eqnarray}
where $e_1 = 1$ and $e_2 = -1$ are the gauge charges of the 
$d_1$ and $d_2$ respectively.
In the DBL[2,2] case, there are both bonding and anti-bonding fields 
present for each species $d_1$ and $d_2$.  The Fermi momenta satisfy 
$k_{F\alpha}^{(0)} + k_{F\alpha}^{(\pi)} = 2\pi\rho$, where $\rho$
is the original boson density per site.
In the DBL[2,1] case, the $d_2$ fermions have only bonding fields.

The allowed interactions contain general density-density terms,
\begin{eqnarray}
\label{h4fermI}
h_{4ferm}^{(I)} = \sum_{b,b', P} 
\bigl[B_{b,b'} \rho_{bP} \rho_{b'P} + C_{b,b'} \rho_{bP} \rho_{b'-P}
\bigr] ~,
\end{eqnarray}
where $b, b'$ sum over all bands of all species, and 
also the following terms,
\begin{eqnarray}
\label{h4fermII}
h_{4ferm}^{(II)} &=& 
\sum_\alpha E_\alpha \left[
d_{\alpha R}^{(0)\dagger} d_{\alpha L}^{(0)\dagger} 
d_{\alpha R}^{(\pi)} d_{\alpha L}^{(\pi)} + \Hc \right] \\
&+& F \left[\sum_P d_{1P}^{(0)\dagger} d_{2-P}^{(0)\dagger} 
                   d_{1-P}^{(\pi)} d_{2P}^{(\pi)} + \Hc \right] \\ 
&+& G \left[\sum_P d_{1P}^{(0)\dagger} d_{2-P}^{(\pi)\dagger} 
                   d_{1-P}^{(\pi)} d_{2P}^{(0)} + \Hc \right].
\end{eqnarray}
In the DBL[2,1] case, all terms that contain $d_2^{(\pi)}$ are
absent.

The above QED2-like theory can be analyzed perturbatively in the 
matter-gauge coupling, e.g., in a systematic $1/N$ expansion, as 
was done in Ref.~\onlinecite{KimLee} for the spinon-gauge treatment 
of the 1D Heisenberg spin chain.  From such studies, we often use the 
following ``Amperean'' rule of thumb:
Gauge interactions modify in a singular way processes that involve 
fermion fields with oppositely oriented group velocities.
If the corresponding gauge currents are parallel (antiparallel),
such processes are enhanced (suppressed), which originates from the 
attraction (repulsion) of currents in electromagnetism.
As an example, $d_{1R}^\dagger d_{2L}^\dagger$ has oppositely charged 
particles moving in opposite directions and producing parallel gauge
currents, so this bilinear is expected to be enhanced compared to the 
mean field.
We will see this explicitly in a solution of the 1D gauge theory by 
bosonization,\cite{Hosotani, KimLee} which we pursue instead of 
the perturbative treatment.
We also caution here and will see below that the different Fermi 
velocities and the general allowed density-density interactions
complicate the analysis significantly and can independently affect 
power laws, in addition to the above Amperean rule for the gauge 
interaction effects.

To bosonize,\cite{Shankar, Lin, Fjaerestad} we write
\begin{equation}
d_{\alpha P}^{(k_y)} = \eta_\alpha^{(k_y)}
\exp[i (\phi_\alpha^{(k_y)} + P \theta_\alpha^{(k_y)})] .
\label{bosonize}
\end{equation}
Here $\phi_\alpha^{(k_y)}$ and $\theta_\alpha^{(k_y)}$ are the
conjugate phase and phonon fields for each band, while
$\eta_\alpha^{(k_y)}$ are Klein factors, which we take to
be Hermitian operators that commute with the bosonic fields and 
anticommute among themselves.
The kinetic Hamiltonian density becomes
\begin{eqnarray}
h_{kinetic} = \sum_{\alpha, k_y} \frac{v_{F\alpha}^{(k_y)}}{2\pi}
\left[ (\partial_x \phi_\alpha^{(k_y)} - e_\alpha a_x)^2 
       + (\partial_x \theta_\alpha^{(k_y)})^2 \right] .
\end{eqnarray}
The density-density interactions Eq.~(\ref{h4fermI}) lead to generic
terms of the form 
$(\partial_x \phi_b - e_b a_x) (\partial_x \phi_{b'} - e_{b'} a_x)$
and $\partial_x \theta_b \partial_x \theta_{b'}$, which are strictly 
marginal and affect power laws.
The other interactions Eq.~(\ref{h4fermII}) produce cosines and
are written separately for the DBL[2,1] and DBL[2,2] below.

We proceed in the Euclidean path integral, choose the gauge $a_x = 0$ 
and integrate out the $a_\tau$.  This renders the field 
$\theta_c \sim \sum_{\alpha, k_y} e_\alpha \theta_\alpha^{(k_y)}$ 
massive and essentially pins it to zero.  This is the suppression of the 
charge fluctuations in the gauge theory and realizes the microscopic 
constraints $d_1^\dagger d_1 = d_2^\dagger d_2$ at long wavelengths.
We then have two gapless modes left in the DBL[2,1] and three modes
in the DBL[2,2], if we can assume that the cosines corresponding to 
$h_{4ferm}^{(II)}$ are irrelevant (see below).

To characterize the phases of the original hard core bosons, 
we examine the single boson correlations, the boson density and 
current correlations, and the pair boson correlations.

The microscopic boson field is written as
\begin{eqnarray}
\label{b}
b(x, y) = d_1(x, y) d_2(x, y)
= \sum_{k_{1y}, k_{2y}, P_1, P_2} \\
e^{i (P_1 k_{F1}^{(k_{1y})} + P_2 k_{F2}^{(k_{2y})}) x} 
e^{i (k_{1y} + k_{2y}) y} 
d_{1 P_1}^{(k_{1y})} d_{2 P_2}^{(k_{2y})},
\nonumber
\end{eqnarray}
which we then express in terms of the continuum bosonized fields
(see separate treatments of the DBL[2,1] and DBL[2,2] below).
For oppositely moving $P_1$ and $P_2$, the $\theta$ part has a non-zero 
projection onto $\theta_c$, and gapping out the $\theta_c$ by the 
gauge fluctuations enhances the corresponding contribution,
in agreement with the Amperean rule applied to the oppositely
charged $d_1$ and $d_2$.

The fermion densities are
\begin{eqnarray}
\label{rhod}
&&\rho_{d_\alpha}(x, y) = d_\alpha^\dagger(x, y) d^\pd_\alpha(x, y) \\
&&= \sum_{P, P', k_y} d_{\alpha P}^{(k_y)\dagger} d_{\alpha P'}^{(k_y)}
          e^{-i (P - P') k_{F\alpha}^{(k_y)} x} 
\nonumber \\
&&+ e^{i\pi y} \sum_{P,P'} \!
\left(  d_{\alpha P}^{(0)\dagger} d_{\alpha P'}^{(\pi)}
      + d_{\alpha -P'}^{(\pi)\dagger} d_{\alpha -P}^{(0)} \right) \!
e^{-i(P k_{F\alpha}^{(0)} - P' k_{F\alpha}^{(\pi)}) x} .
\nonumber
\end{eqnarray}
The two lines in the expansion separate $q_y=0$ and $q_y=\pi$ parts,
and all contributions with the same non-zero momentum are grouped
together (assuming generic distinct bonding and antibonding bands).
We also consider the particle current on the rungs 
(the definition below uses the gauge choice $a_y=0$ made early on):
\begin{eqnarray}
\label{jrung}
&&\!\!\!
j_{\perp, \alpha}(x) =
\frac{i}{2}\left[  d_\alpha^\dagger(x, 1) d^\pd_\alpha(x, 2) 
                 - d_\alpha^\dagger(x, 2) d^\pd_\alpha(x, 1) \right] \\
&&= i \sum_{P,P'} \!
\left(  d_{\alpha P}^{(0)\dagger} d_{\alpha P'}^{(\pi)}
      - d_{\alpha -P'}^{(\pi)\dagger} d_{\alpha -P}^{(0)} \right) \!
e^{-i(P k_{F\alpha}^{(0)} - P' k_{F\alpha}^{(\pi)}) x} .
\nonumber
\end{eqnarray}
The continuum expansion resembles that for the density at $q_y=\pi$, 
but with opposite signs between 
$d_{\alpha P}^{(0)\dagger} d_{\alpha P'}^{(\pi)}$
and $d_{\alpha -P'}^{(\pi)\dagger} d_{\alpha -P}^{(0)}$; 
this will help to distinguish different phases in the analysis of 
instabilities, Secs.~\ref{subapp:DBL21instab},~\ref{subapp:DBL22instab}.
Bosonized expressions for the contributing bilinears will be
given separately in the DBL[2,1] and DBL[2,2] cases.
Here we only note that when the constituent particle and hole are on the 
opposite sides, we again have non-zero projection onto $\theta_c$, 
so the corresponding contribution is enhanced by the gauge fluctuations, 
in agreement with the Amperean rule.

Finally, we consider the boson pair operator 
$b(x_1, y_1) b(x_2, y_2)$ and expand in terms of the 
continuum fermion fields.  Assuming $x_1$ and $x_2$ are nearby, 
we characterize the pair by its center of mass coordinate 
$X = (x_1+x_2)/2$ along the chains (this becomes the argument of the 
continuum fields) and also by the internal pair structure.  To this end,
we collect different microscopic contributions that give rise to each 
term  $d_{1a}(X) d_{1b}(X) d_{2c}(X) d_{2d}(X)$, 
where $a,b,c,d$ are combined band and Fermi point indices.
Using short-hands
$D_{1a}({\bf r}) = d_{1a} \exp[i {\bf k}_{1a} \cdot {\bf r}]$,
etc., then $b(x_1, y_1) b(x_2, y_2)$ contains
\begin{eqnarray*}
&-&\left[  D_{1a}(x_1, y_1) D_{1b}(x_2, y_2)
         - D_{1a}(x_2, y_2) D_{1b}(x_1, y_1)\right] \\
&&\times
   \left[  D_{2c}(x_1, y_1) D_{2d}(x_2, y_2)
         - D_{2c}(x_2, y_2) D_{2d}(x_1, y_1)\right] \\
&=& d_{1a}(X) d_{1b}(X) d_{2c}(X) d_{2d}(X) \\
&&\times e^{i(k_{1ax} + k_{1bx} + k_{2cx} + k_{2dx}) X}
\Phi_{1a, 1b, 2c, 2d}(y_1, y_2, \xi) ~,
\end{eqnarray*}
where $\xi = x_1 - x_2$ is the relative coordinate along the chains.
The ``pair wavefunction'' is
\begin{eqnarray}
\label{Phipair}
\Phi_{1a, 1b, 2c, 2d}(y_1, y_2, \xi) = 
e^{i(k_{1ay} + k_{1by} + k_{2cy} + k_{2dy}) Y}\\
\times 4 
\sin\left( \frac{{\bf k}_{1a} - {\bf k}_{1b}}{2} \cdot {\bm \rho} \right)
\sin\left( \frac{{\bf k}_{2c} - {\bf k}_{2d}}{2} \cdot {\bm \rho} \right)
 ~,
\nonumber
\end{eqnarray}
with $Y=\frac{y_1+y_2}{2}$ and ${\bm \rho} = (\xi, y_1 - y_2)$.
In the above, we have chosen to characterize the pair by the momentum 
along the $x$-axis, but not by the momentum along the $y$-axis.  
On the 2-leg ladder, it is easier to visualize pairs by keeping track
of both $y_1$ and $y_2$.

A few more words about this characterization on the 2-leg ladder.
Consider first a 2D setting, where we would write a contribution 
to a boson-pair $b({\bf r}_1) b({\bf r}_2)$ by say
\begin{equation}
\hat{O}({\bf R}) e^{i {\bf q} \cdot {\bf R}} \Phi({\bm \rho}) ~,
\label{Phipair2D}
\end{equation}
where $\hat{O}$ is some slowly varying operator of the center of
mass coordinate ${\bf R}$, while $\Phi({\bm \rho})$ is the internal 
pair function in the relative coordinate ${\bm \rho}$.  
Specializing to a square lattice, we could then distinguish s-wave and 
d-wave (more precisely, $d_{xy}$) pairing by looking at the signs of 
$\Phi({\bm \rho} = \hat{x} \pm \hat{y})$.
On the 2-leg ladder, however, separating out the transverse momentum
$q_y$ as in 2D mixes things a little.
Specifically, a rotation of an $\hat{x} + \hat{y}$ pair into an 
$\hat{x} - \hat{y}$ pair can be also achieved by a ``translation''
$y \to y+1 \mod 2$.  To avoid this, we instead define the
pair-function by writing
\begin{equation}
\hat{O}_{\rm 2leg} (X) e^{i q_x X} \Phi_{\rm 2leg}(\xi, y_1, y_2) ~,
\end{equation}
where $X$ and $\xi$ are the $x$-components of ${\bf R}$ and ${\bm \rho}$
respectively.  We then characterize the pair wavefunction
$\Phi_{\rm 2leg}$ by the symmetry under interchanging the two legs
of the ladder ($y_{1,2} \to y_{1,2} + 1 \mod 2$).
When $\Phi_{\rm 2leg}$ is even, we call it s-wave, since then the
amplitudes for a pair sitting on a $+45^\circ$ diagonal and 
a pair on a $-45^\circ$ diagonal are the same.
When $\Phi_{\rm 2leg}$ is odd, we call it d-wave, since then the 
amplitudes for the two diagonals are opposite.  The names ``s-wave'' and
``d-wave'' are used in anticipation that it is such local pictures
of the pairs on the plaquettes that survive when we build towards 2D.
Returning to the DBL construction Eq.~(\ref{Phipair}),
we get s-wave or d-wave character depending whether 
$k_{1ay} + k_{1by} + k_{2cy} + k_{2dy}$ is even or odd multiple
of $\pi$.

We now specialize to the DBL[2,1] and DBL[2,2] cases in turn.

\subsection{DBL[2,1]}
\label{subapp:DBL21}
Upon bosonization, we start with three modes
$(\phi_1^{(0)}, \theta_1^{(0)})$, $(\phi_1^{(\pi)}, \theta_1^{(\pi)})$,
and $(\phi_2, \theta_2)$ (we drop the bonding band label for the $d_2$ 
fermions).  
To proceed formally, we change to new canonical variables via
\begin{eqnarray*}
&\phi_\rho = \frac{1}{2} (\phi_1^{(0)} + \phi_1^{(\pi)} + 2 \phi_2),\quad
&\theta_\rho = \frac{1}{3} (\theta_1^{(0)} + \theta_1^{(\pi)} + 2 \theta_2); \\
&\phi_c = \frac{1}{2} (\phi_1^{(0)} + \phi_1^{(\pi)} - \phi_2), \quad
&\theta_c = \frac{2}{3} (\theta_1^{(0)} + \theta_1^{(\pi)} - \theta_2);\\
&\phi_{-} = \frac{1}{2} (\phi_1^{(0)} - \phi_1^{(\pi)}), \quad
&\theta_{-} = \theta_1^{(0)} - \theta_1^{(\pi)}.
\end{eqnarray*}

The microscopic boson field $b({\bf r})$ has the following contributions 
listed by their momentum (cf.~Eq.~\ref{b}):
\begin{eqnarray*}
& (k_{F1}^{(k_y)} + k_{F2}, k_y) : \;
& d_{1R}^{(k_y)} d_{2R} 
\sim e^{i (\phi_\rho \pm \phi_{-})}
     e^{i (\frac{3}{2} \theta_\rho \pm \frac{1}{2} \theta_{-})} ~; \\
& (k_{F1}^{(k_y)} - k_{F2}, k_y) : \;
& d_{1R}^{(k_y)} d_{2L} 
\sim e^{i (\phi_\rho \pm \phi_{-})}
     e^{i (\theta_c - \frac{1}{2} \theta_\rho 
           \pm \frac{1}{2} \theta_{-})} ~.
\end{eqnarray*}
In each equation, the top and bottom signs are for $k_y = 0$ and $\pi$
respectively.  We omit the Klein factors for simplicity since we 
will not need them here.  Also, we do not show combinations that can be 
obtained from the above by reversing all momenta.

The microscopic boson density has contributions from both the 
$d_1$ and $d_2$ particles via Eq.~(\ref{rhod}), 
while the rung current has contributions only from the $d_1$ 
via Eq.~(\ref{jrung}); we list all involved bilinears by their
momentum:
\begin{eqnarray*}
&(0, 0) : \;
&\sum_{k_y, P} d_{1P}^{(k_y)\dagger} d_{1P}^{(k_y)}
= \frac{\partial_x (\theta_1^{(0)} + \theta_1^{(\pi)})}{\pi} ~; \\
&(2 k_{F1}^{(k_y)}, 0) : \;
&d_{1R}^{(k_y)\dagger} d_{1L}^{(k_y)} 
= e^{-i (\theta_c + \theta_\rho \pm \theta_{-})} ~; \\
&(k_{F1}^{(0)} - k_{F1}^{(\pi)}, \pi) : \;
&d_{1R}^{(0)\dagger} d_{1R}^{(\pi)} 
= \eta_1^{(0)} \eta_1^{(\pi)} e^{-i 2 \phi_{-}} e^{-i \theta_{-}} ~; \\
&
&d_{1L}^{(\pi)\dagger} d_{1L}^{(0)}
= \eta_1^{(\pi)} \eta_1^{(0)} e^{i 2 \phi_{-}} e^{-i \theta_{-}} ~; \\
&(k_{F1}^{(0)} + k_{F1}^{(\pi)}, \pi) : \;
&d_{1R}^{(0)\dagger} d_{1L}^{(\pi)}
= \eta_1^{(0)} \eta_1^{(\pi)} e^{-i 2 \phi_{-}} e^{-i (\theta_c + \theta_\rho)} ~; \\
&
&d_{1R}^{(\pi)\dagger} d_{1L}^{(0)}
= \eta_1^{(\pi)} \eta_1^{(0)} e^{i 2 \phi_{-}} e^{-i (\theta_c + \theta_\rho)} ~; \\
&(0, 0) : \;
&\sum_P d_{2P}^{\dagger} d_{2P} = \frac{\partial_x \theta_2}{\pi}~;\\ 
&(2 k_{F2}, 0) : \;
&d_{2R}^\dagger d^\pd_{2L} = e^{i (\theta_c - 2 \theta_\rho)} ~.
\end{eqnarray*}
Here we include the Klein factors, but do not show combinations
that can be obtained from the above by reversing all momenta.

When we include gauge fluctuations, $\theta_c$ becomes massive,
locking together $\theta_1^{(0)} + \theta_1^{(\pi)}$ and $\theta_2$.
Integrating out $(\phi_c, \theta_c)$, we obtain a generic 
harmonic liquid theory in terms of two coupled modes 
$(\phi_\rho, \theta_\rho)$ and $(\phi_{-}, \theta_{-})$.
Of the cosine interactions Eq.~(\ref{h4fermII}), we have only
\begin{equation}
h_{4ferm}^{II} = 2 E_1 \cos(4\phi_{-}) ~.
\label{4fermE1}
\end{equation}
Assuming this is irrelevant, we have a stable phase with two gapless 
modes.  The potential instability due to this term is considered in
Sec.~\ref{subapp:DBL21instab} below.

We do not write explicitly the full DBL[2,1] theory in terms of the
$(\phi_\rho, \theta_\rho)$ and $(\phi_{-}, \theta_{-})$ since even 
without the additional interactions from Eq.~(\ref{h4fermI}) we 
have a general coupled harmonic system.
We can still make some observations about the scaling dimensions 
of the contributions to the boson and the boson density or current 
operators.
In the initial fermionic mean field, the bands $d_1^{(0/\pi)}, d_2$
are decoupled and the scaling dimensions of all contributions to the 
operators $b$ and $\hat\rho_b$ are equal to 1.  
The gauge fluctuations effectively set $\theta_c = 0$ everywhere.

Consider first the contributions to $b$.
For arbitrary couplings in the full harmonic theory, 
we can argue that the scaling dimensions of the operators 
$d_{1R}^{(k_y)} d_{2R}$ are at least $1$,
i.e.~larger than the mean field value (these have no explicit Amperean 
enhancement).
On the other hand, setting $\theta_c = 0$, the scaling dimensions of the 
operators $d_{1R}^{(k_y)} d_{2L}$ can be lower than
the mean field in accord with the Amperean rule;
without further information, we cannot say much -- in general,
these scaling dimensions can be as low as zero.

Consider now the contributions to the boson density $\hat\rho_b$.
The zero momentum contribution has scaling dimension $1$.
We can argue for arbitrary couplings in the theory that the 
$(k_{F1}^{(0)} - k_{F1}^{(\pi)}, \pi)$ contribution has scaling 
dimension larger than the mean field value $1$.
On the other hand, the $(2k_{F1}^{(k_y)}, 0)$, 
$(k_{F1}^{(0)} + k_{F1}^{(\pi)}, \pi)$, and $(2k_{F2}, 0)$
contributions can have smaller scaling 
dimensions upon setting $\theta_c = 0$, in accord with the Amperean
enhancement rule.  We do not have general bounds on the scaling 
dimensions in the coupled two-mode system, although clearly these are 
not all independent.
Furthermore, in all of the preceding discussion, even if the gauge
interaction acts to enhance some correlation, the short-range 
interactions in 1D can act to suppress the correlation, so the ultimate 
fate is not clear.

As an illustration, we list all scaling dimensions in the case 
when the $(\phi_\rho, \theta_\rho)$ and $(\phi_{-}, \theta_{-})$ 
modes are decoupled and characterized by the Luttinger parameters 
$g_\rho$ and $g_{-}$ respectively
[our convention is that $g$ enters as
$g (\partial_x \phi)^2 + \frac{1}{g} (\partial_x \theta)^2$
in the action]:
\begin{eqnarray*}
\Delta[b_{(k_{F1}^{(k_y)} + k_{F2}, k_y)}]
 &=& \frac{3}{8} (\frac{2}{3 g_\rho} + \frac{3 g_\rho}{2})
   + \frac{1}{8} (\frac{2}{g_{-}} + \frac{g_{-}}{2}) \geq 1 ~; \\
\Delta[b_{(k_{F1}^{(k_y)} - k_{F2}, k_y)}]
 &=& \frac{1}{8} (\frac{2}{g_\rho} + \frac{g_\rho}{2})
   + \frac{1}{8} (\frac{2}{g_{-}} + \frac{g_{-}}{2}) \geq 1/2 ~; \\
\Delta[\rho_{b, (2k_{F1}^{(k_y)}, 0)}] 
 &=& \frac{1}{4} g_\rho + \frac{1}{4} g_{-} ~; \\
\Delta[\rho_{b, (k_{F1}^{(0)} - k_{F1}^{(\pi)}, \pi)}] 
 &=& \frac{1}{2} (\frac{2}{g_{-}} + \frac{g_{-}}{2}) \geq 1 ~; \\
\Delta[\rho_{b, (k_{F1}^{(0)} + k_{F1}^{(\pi)}, \pi)}] 
 &=& \frac{1}{g_{-}} + \frac{1}{4} g_\rho~; \\
\Delta[\rho_{b, (2k_{F2}, 0)}] &=& g_\rho ~.
\end{eqnarray*}
This example is loosely motivated by the observation in
Fig.~\ref{fig:dbl_nq} that the boson $n({\bf q})$ has roughly similar 
singularities at $(k_{F1}^{(0)} - k_{F2}, 0)$ and 
$(k_{F1}^{(\pi)} - k_{F2}, \pi)$.  Also, increasing the ring term $K$ 
and decreasing the interchain hopping $J_\perp$ drives the mean 
field bonding and antibonding bands to be more similar, suggesting
such approximate decoupling of the two modes.
Continuing with this illustration, the boson correlators at 
$(k_{F1}^{(k_y)} - k_{F2}, k_y)$ are most enhanced when
$g_\rho=2$ and $g_{-}=2$ giving
$\Delta[b_{(k_{F1}^{(k_y)} - k_{F2}, k_y)}] = 1/2$;
with such couplings we also find mean-field-like 
$\Delta[\rho_{b, (2k_{F1}^{(k_y)}, 0)}]
= \Delta[\rho_{b, (k_{F1}^{(0)} - k_{F1}^{(\pi)}, \pi)}] 
= \Delta[\rho_{b, (k_{F1}^{(0)} + k_{F1}^{(\pi)}, \pi)}] = 1$,
while $\Delta[\rho_{b, (2k_{F2}, 0)}] = 2$.
These numbers are qualitatively similar to the singularities detectable
by eye in the density structure factor in Fig.~\ref{fig:dbl_nq},
where we see some signatures at $(2k_{F1}^{(0/\pi)}, 0)$,
$(k_{F1}^{(0)} \pm k_{F1}^{(\pi)}, \pi)$, and no visible 
signature at $(2k_{F2}, 0)$.
A more careful look at the DMRG data shows that the density singularities
at $(2k_{F1}^{(0/\pi)}, 0)$ are slightly stronger than the mean field,
while the ones at $(k_{F1}^{(0)} \pm k_{F1}^{(\pi)}, \pi)$ are weaker;
this would suggest $g_{-}$ somewhat smaller than $2$, which would also
be consistent with the assumed irrelevance of the cosine interaction 
Eq.~(\ref{4fermE1}).  However the DMRG boson correlator singularity 
has estimated scaling dimension somewhat smaller than the maximum $1/2$ 
in this illustration, so one should not take the above too literally.
Nevertheless, this example gives some sense to the strengths of the 
observed singularities in Fig.~\ref{fig:dbl_nq}.

Finally, consider a boson pair operator; we list contributing 
four-fermion combinations by their momentum along the chains:
\begin{eqnarray*}
&0 : \;
&d_{1R}^{(k_y)} d_{1L}^{(k_y)} d_{2R} d_{2L}
\sim e^{i (2 \phi_\rho \pm 2 \phi_{-})} ~; \\
&P (k_{F1}^{(0)} + k_{F1}^{(\pi)}) : \;
&d_{1P}^{(0)} d_{1P}^{(\pi)} d_{2R} d_{2L}
\sim e^{i 2 \phi_\rho} e^{i P (\theta_c + \theta_\rho) } ~; \\
&P (k_{F1}^{(0)} - k_{F1}^{(\pi)}) : \;
&d_{1P}^{(0)} d_{1-P}^{(\pi)} d_{2R} d_{2L}
\sim e^{i 2 \phi_\rho} e^{i P \theta_{-}} ~.
\end{eqnarray*}
As explained earlier, we characterize each contribution by the
internal pair wavefunction Eq.~(\ref{Phipair}).
Again we have omitted the Klein factors but can easily restore them
when needed.

For the zero momentum contribution we have
\begin{eqnarray}
\label{Phiq0}
\Phi_{1R(k_y), 1L(k_y), 2R, 2L}(y_1, y_2, \xi) = \\
= 4 e^{i k_y (y_1 + y_2)} \sin(k_{F1}^{(k_y)} \xi) \sin(k_{F2} \xi) ~.
\nonumber
\end{eqnarray}
This remains unchanged if we simultaneously change the $y$ coordinates
of both bosons in the pair, e.g., if we move a pair that lies entirely
in chain 1 to chain 2 or if we turn a $+45^\circ$ diagonal pair into 
a $-45^\circ$ diagonal pair; we call this s-wave.

For the contribution at momentum 
$k_{F1}^{(0)} + k_{F1}^{(\pi)} = 2\pi\rho$ we have
\begin{eqnarray}
\label{Phiq2pirho}
&&\Phi_{1R(0), 1R(\pi), 2R, 2L}(y_1, y_2, \xi) = \\
&&= 4 \sin(k_{F2} \xi)  e^{i\pi \frac{y_1+y_2}{2}} 
\sin\left[ \frac{k_{F1}^{(0)} - k_{F1}^{(\pi)}}{2} \xi 
           - \frac{\pi}{2} (y_1 - y_2) \right]
\nonumber \\
&&= 4 \sin(k_{F2} \xi) 
\biggl( \delta_{y_1 = y_2} (-1)^{y_1} 
        \sin\left[\frac{k_{F1}^{(0)} - k_{F1}^{(\pi)}}{2} \xi \right] 
\nonumber \\
&&\quad\quad\quad\quad\quad
       + i \delta_{y_1 \neq y_2} (-1)^{y_1}
         \cos\left[\frac{k_{F1}^{(0)} - k_{F1}^{(\pi)}}{2} \xi \right]
\biggr) ~.
\nonumber
\end{eqnarray}
This changes sign if we simultaneously change both $y$ coordinates
in the pair.  When $k_{F1}^{(0)}$ and $k_{F1}^{(\pi)}$ approach 
each other, the pairs straddling the two chains have larger amplitude.
Since the amplitude changes sign when we turn a $+45^\circ$ 
diagonal pair into a $-45^\circ$ diagonal pair, the internal
structure of the pair has some d-wave character.
The contribution at $k_{F1}^{(0)} - k_{F1}^{(\pi)}$ can be similarly 
characterized, but is omitted here.

In the free fermion mean field all contributions have scaling 
dimension 2.  Beyond the mean field, the d-wave contribution at 
$k_{F1}^{(0)} + k_{F1}^{(\pi)}$ is potentially enhanced by setting 
$\theta_c = 0$; still, its scaling dimension is at least $1$, 
while we cannot say much about the scaling dimension of the 
s-wave contribution at zero momentum.
In the above illustration with decoupled $\phi_\rho$ and $\phi_-$ modes, 
the d-wave pair has scaling dimension $1/g_\rho + g_\rho/4 = 1$ if
we set $g_\rho = 2$, while the s-wave pair has scaling dimension
is $1/g_\rho + 1/g_{-} = 1$ if we also set $g_{-} = 2$.
The latter would be larger if we take smaller $g_{-}$, in agreement
with the observed dominance of the d-wave-like pair correlations
in the DMRG Fig.~\ref{fig:dbl_pair} oscillating at wavevector $2\pi\rho$.

\subsection{Possible Instability of the DBL[2,1]}
\label{subapp:DBL21instab}

Let us ask what phase is obtained starting from the DBL[2,1] if
the interaction Eq.~(\ref{4fermE1}) is relevant
[the Amperean rule does not point one way or another, but additional 
interactions from Eq.~(\ref{h4fermI}) can make this happen].
Then $\phi_{-}$ is pinned, with two distinct cases depending
on the sign of $E_1$ considered below, while $\theta_{-}$ fluctuates 
wildly.  All contributions to $b$ contain $\theta_{-}$ in the 
exponent, so the single boson correlation function decays exponentially
-- there is a charge gap to single boson excitations.
The $(\phi_\rho, \theta_\rho)$ mode still remains gapless, and we have 
a single-mode harmonic fluid described by the parameter $g_\rho$.  

The boson density shows power law correlations at zero momentum 
with scaling dimension 1.
The remaining particle-hole bilinears with power law correlations are
at wavevector $(k_{F1}^{(0)} + k_{F1}^{(\pi)}, \pi) = (2\pi\rho, \pi)$ 
(scaling dimension $g_\rho/4$) and 
at $(2 k_{F2}, 0) = (4\pi\rho, 0)$ (scaling dimension $g_\rho$).
The ${\bf q} = {\bf 0}$ and ${\bf q} = (2 k_{F2}, 0)$ parts do not 
distinguish qualitatively between the two possibilities $E_1 > 0$ and 
$E_1 < 0$, while the ${\bf q} = (k_{F1}^{(0)} + k_{F1}^{(\pi)}, \pi)$ 
part does distinguish.
Explicitly, the $d_1$ particle density is
\begin{eqnarray*}
\rho_{d_1}(x,y) \!=\! 
- 2 i \eta_1^{(0)} \eta_1^{(\pi)} \sin(2 \phi_{-}) e^{i \theta_\rho}
e^{i(k_{F1}^{(0)} + k_{F1}^{(\pi)}) x + i\pi y}
\!+\! \Hc,
\end{eqnarray*}
where we have omitted the zero momentum component and parts with
exponentially vanishing correlations.
On the other hand, the $d_1$ particle rung current is
\begin{eqnarray*}
j_{\perp,1}(x) \!=\!
2 i \eta_1^{(0)} \eta_1^{(\pi)} \cos(2 \phi_{-}) e^{i \theta_\rho}
e^{i(k_{F1}^{(0)} + k_{F1}^{(\pi)}) x}
+ \Hc ~.
\end{eqnarray*}

Consider now two cases:

\underline{$E_1 > 0$}:  It follows that 
$2\phi_{-} = \frac{\pi}{2} \mod \pi$.  In this case $\rho_{d_1}$ shows
power law correlations at $(k_{F1}^{(0)} + k_{F1}^{(\pi)}, \pi)$,
while $j_{\perp,1}$ correlations are absent.

\underline{$E_1 < 0$}:  It follows that
$2\phi_{-} = 0 \mod \pi$.  In this case $\rho_{d_1}$ correlations at the
above wavevector are absent, while $j_{\perp,1}$ shows power law
at wavevector $k_{F1}^{(0)} + k_{F1}^{(\pi)}$ along the chains.

Let us finally consider the pair-boson correlations, which are also
power law.  Putting together all contributions
to $b(x_1,y_1) b(x_2,y_2)$, we have 
\begin{eqnarray*}
e^{i 2 \phi_\rho}[e^{i 2 \phi_{-}} \Phi_{1R(0), 1L(0), 2R, 2L}
                     + e^{-i 2 \phi_{-}} \Phi_{1R(\pi), 1L(\pi), 2R, 2L}]
\\
+\sum_P \eta_1^{(0)} \eta_1^{(\pi)} 
e^{i (2 \phi_\rho + P [\theta_\rho + (k_{F1}^{(0)} + k_{F1}^{(\pi)}) X] )} 
\Phi_{1P(0), 1P(\pi), 2R, 2L} ~,
\end{eqnarray*}
where the appropriate $\Phi(y_1, y_2, \xi)$ are given in
Eqs.~(\ref{Phiq0}) and (\ref{Phiq2pirho}).
The pairs at zero momentum have dominant correlations with
scaling dimension $1/g_\rho$; these have s-wave character as far as
rotating diagonal bonds is concerned, but the amplitude details 
depend on the sign of $E_1$.
The pairs at $k_{F1}^{(0)} + k_{F1}^{(\pi)}$ have subdominant
correlations with scaling dimension $1/g_\rho + g_\rho/4$;
these have d-wave character for rotating the diagonal bonds
and the details do not depend much on the sign of $E_1$.

Summarizing, the resulting phase has a gap to single boson excitations, 
but has power law correlations for particle-hole and particle-particle 
composites. 
The dominant particle-hole composite has scaling dimension $g_\rho/4$ 
and either contributes to the density at wavevector $(2\pi\rho, \pi)$ if 
$E_1 > 0$ or to the rung current at $2\pi\rho$ if $E_1 < 0$.
The dominant particle-particle composite has scaling dimension $1/g_\rho$
and represents s-wave pairing at zero momentum.
Loosely speaking, we can describe this phase by saying that the 
bosons form s-wave-like pairs and these molecules in turn form a 
Luttinger liquid.   
The current and/or density fluctuations occur at a wavelength equal
to the mean inter-pair spacing along the ladder, as expected for 
a 1D Luttinger liquid (of pairs).  
The form of the particle-hole fluctuations across the rungs
presumably reflects the internal pair structure in the two cases.

\subsection{DBL[2,2] and Instability Towards S-Wave Pairing}
\label{subapp:DBL22instab}

In the DBL[2,2] case, we start with four bands, 
$d_1^{(0/\pi)}, d_2^{(0/\pi)}$, and have four modes upon
bosonization Eq.~(\ref{bosonize}).
It is convenient to introduce the following canonical variables:
\begin{eqnarray}
\phi_\rho &=& \frac{1}{2} (\phi_1^{(0)} + \phi_1^{(\pi)}
                           + \phi_2^{(0)} + \phi_2^{(\pi)}) ~; \\
\phi_c &=& \frac{1}{2} (\phi_1^{(0)} + \phi_1^{(\pi)}
                        - \phi_2^{(0)} - \phi_2^{(\pi)}) ~; \\
\phi_+ &=& \frac{1}{2} (\phi_1^{(0)} - \phi_1^{(\pi)}
                        + \phi_2^{(0)} - \phi_2^{(\pi)}) ~; \\
\phi_- &=& \frac{1}{2} (\phi_1^{(0)} - \phi_1^{(\pi)}
                        - \phi_2^{(0)} + \phi_2^{(\pi)}) ~;
\end{eqnarray}
with the same transformation for the $\theta$ variables.
The cosine interactions Eq.~(\ref{h4fermII}) are
\begin{eqnarray*}
h_{4ferm}^{(II)} =
2 E_1 \cos(2\phi_{+} + 2\phi_{-}) + 2 E_2 \cos(2\phi_{+} - 2\phi_{-}) \\
- 4 \hat\Gamma F \cos(2\phi_{+}) \cos(2 \theta_c)
+ 4 \hat\Gamma G \cos(2\phi_{-}) \cos(2 \theta_c) ~,
\label{hintIIbos}
\end{eqnarray*}
where 
$\hat\Gamma = \eta_1^{(0)} \eta_1^{(\pi)} \eta_2^{(0)} \eta_2^{(\pi)}$.

Proceeding with the analysis as in the DBL[2,1] case, upon integrating 
out the gauge field, $\theta_c$ becomes massive.
If by tweaking the strictly marginal interactions $h_{4ferm}^{(I)}$ 
we could render the $h_{4ferm}^{(II)}$ terms irrelevant, we would 
end up with a phase with three gapless modes.  We call this possible 
phase DBL[2,2], and it can be analyzed similarly to DBL[2,1].  
For example, the mean field boson correlation in this phase would read
\begin{eqnarray*}
G_b^{MF}(x,y) \sim \sum_{k_{y1}, k_{y2}} 
\frac{e^{i(k_{y1} + k_{y2}) y}}{x^2} 
\Bigl(\cos[(k_{F1}^{(k_{y1})} - k_{F2}^{(k_{y2})}) x] \\
      - \cos[(k_{F1}^{(k_{y1})} + k_{F2}^{(k_{y2})}) x] \Bigr) ~.
\end{eqnarray*}
The first term in the brackets comes from 
$d_{1P}^{(k_{y1})} d_{2-P}^{(k_{y2})}$ and is expected
to be enhanced by Amperean attraction (in the bosonization, this
term contains $\theta_c$ in the exponent and is potentially enhanced 
upon setting $\theta_c=0$).
On the other hand, the second term in the brackets has no
Amperean enhancement.  
As emphasized earlier, besides the gauge fluctuations crudely captured 
by the Amperean rule, other interactions can also change the scaling 
dimensions.
Interestingly, and this has important consequences explored below,
$d_{1R}^{(0)} d_{2L}^{(0)}$ and $d_{1L}^{(\pi)} d_{2R}^{(\pi)}$ carry the
same momentum 
$(k_{F1}^{(0)} - k_{F2}^{(0)}, 0)
 = (k_{F2}^{(\pi)} - k_{F1}^{(\pi)}, 0)$, 
and also $d_{1R}^{(0)} d_{2L}^{(\pi)}$ and $d_{1L}^{(\pi)} d_{2R}^{(0)}$ 
carry the same momentum 
$(k_{F1}^{(0)} - k_{F2}^{(\pi)}, \pi)
= (k_{F2}^{(0)} - k_{F1}^{(\pi)}, \pi)$.
In fact, Josephson-like coupling between the first two ``bosonic modes'' 
is allowed and gives rise to the $F$ interaction in Eq.~(\ref{h4fermII}),
while Josephson coupling between the last two modes gives rise to the 
$G$ interaction.  If the gauge physics is the dominant determining factor
in the DBL[2,2] phase, we expect the state to be strongly unstable to 
these interactions, and we pursue this scenario below.  
On the other hand, 
if the DBL[2,2] were stabilized against the $F$ and $G$ by some 
density-density interactions dominating over the gauge physics,
which is possible and interesting, we would have little intuition
from the gauge theory perspective, and we do not pursue this possibility 
further.

If the gauge physics dominates, it appears very likely that
the $F$ and $G$ terms are relevant, as can be seen from their
bosonized expressions upon setting $\theta_c = 0$ 
(in accord with the Amperean rule).
Let us explore the resulting phase when $F$ and $G$ flow to
large values and pin the fields $\phi_{+}$ and $\phi_{-}$.
Four different possibilities depending on the signs of $F$ and $G$ are 
discussed below.  In such a phase, the conjugate fields $\theta_{+}$ and 
$\theta_{-}$ fluctuate strongly and any operator containing these in the 
exponent will have only short-range correlations.
Thus, we conclude that the boson correlations decay exponentially, i.e., 
there is a gap to single boson excitations.  However, there is still one 
gapless mode $(\phi_\rho, \theta_\rho)$, which we characterize with 
the Luttinger parameter $g_\rho$.

Consider fermion bilinears composed of a particle and a hole of the 
same species that enter density and current,
Eqs.~(\ref{rhod},\ref{jrung}).  
Other than the $q=0$ densities 
$d_{\alpha P}^{(k_y)\dagger} d_{\alpha P}^{(k_y)}$,
the following bilinears survive when $\phi_\pm$ get pinned:
\begin{eqnarray}
&&\!\!\!\!\!\!\!\!\!\!\!\!
d_{\alpha P}^{(0)\dagger} d_{\alpha -P}^{(\pi)}
= \eta_\alpha^{(0)} \eta_\alpha^{(\pi)}
e^{-i (\phi_{+} + e_\alpha \phi_{-})} 
e^{-i P (\theta_\rho + e_\alpha \theta_c)}, \\
&&\!\!\!\!\!\!\!\!\!\!\!\!
d_{\alpha P}^{(\pi)\dagger} d_{\alpha -P}^{(0)} 
= \eta_\alpha^{(\pi)} \eta_\alpha^{(0)}
e^{i (\phi_{+} + e_\alpha \phi_{-})} 
e^{-i P (\theta_\rho + e_\alpha \theta_c)}.
\end{eqnarray}
The particle density Eq.~(\ref{rhod}) is, upon dropping the
zero momentum part and setting $\theta_c=0$,
\begin{eqnarray}
\rho_{d_\alpha}(x,y) =
-2 i \eta_\alpha^{(0)} \eta_\alpha^{(\pi)}
\sin(\phi_{+} + e_\alpha \phi_{-}) \\
\times e^{i \pi y}
\sum_P e^{i P \theta_\rho} 
       e^{i P (k_{F\alpha}^{(0)} + k_{F\alpha}^{(\pi)})x} ~.
\nonumber
\end{eqnarray}
The particle current on the rungs Eq.~(\ref{jrung}) is
\begin{eqnarray}
j_{\perp,\alpha}(x) =
2 i \eta_\alpha^{(0)} \eta_\alpha^{(\pi)} 
\cos(\phi_{+} + e_\alpha \phi_{-}) \\
\times \sum_P e^{i P \theta_\rho}
              e^{i P (k_{F\alpha}^{(0)} + k_{F\alpha}^{(\pi)})x} ~.
\nonumber
\end{eqnarray}

We now discuss different phases that can arise depending on the
signs of the couplings $F$ and $G$.  
We set $\Gamma = 1$, from which it follows
$\la \eta_2^{(0)} \eta_2^{(\pi)} \ra = 
-\la \eta_1^{(0)} \eta_1^{(\pi)} \ra$
(the physical results are independent of the choice of $\Gamma$).
There are four cases:

\underline{$F>0, G>0$}:
It follows that $\phi_{+} = 0 \mod \pi$, 
$\phi_{-} = \frac{\pi}{2} \mod \pi$.  
In this case $\rho_{d_1} = \rho_{d_2}$ show power law correlations at 
$(2\pi\rho, \pi)$ with the scaling dimension $g_\rho/4$, 
while $j_{\perp,\alpha}$ correlations are absent.
This is a natural phase coming from the microscopic gauge theory since
the fluctuations of the $d_1$ and $d_2$ densities are in sync;
also, when we integrate out the massive gauge field $a_I - a_{II}$
early in the derivation of the continuum theory, we in fact generate 
such $F = G > 0$ couplings.
Foretelling the analysis of the pairing correlations below,
we propose that this is the s-wave-paired phase observed in the
DMRG, Sec.~\ref{subsec:s-paired}.

\underline{$F<0, G<0$}:
It follows that $\phi_{+} = \frac{\pi}{2} \mod \pi$,
$\phi_{-} = 0 \mod \pi$.
In this case $\rho_{d_1} = -\rho_{d_2}$ show power law correlations
at $(2\pi\rho, \pi)$, while $j_{\perp,\alpha}$ correlations are absent.
The fluctuations of the $d_1$ and $d_2$ densities are out of sync.
This phase with spatial modulation of the gauge charge is less natural 
coming from the lattice theory.  It is in principle allowed if the 
bare couplings are finite, since then such modulation costs only 
finite energy density that can be offset by some other short-range 
interactions; but the cost is large if the microscopic theory is at 
strong coupling as is usually the case in the slave particle treatments.
For example, only the phase with $\rho_{d_1} = \rho_{d_2}$ can
be realized by our wavefunctions.

\underline{$F<0, G>0$}:
It follows that $\phi_{+} = \phi_{-} = \frac{\pi}{2} \mod \pi$.
In this case $\rho_{d_\alpha}$ correlations are absent, while
$j_{\perp, 1} = j_{\perp, 2}$ show power low correlations 
at wavevector $2\pi\rho$ along the chains with scaling dimension 
$g_\rho/4$.   The $d_1$ and $d_2$ particle currents are in sync,
so this phase is natural in the gauge theory; 
for the original bosons, it would have enhanced current-current 
correlations but not density correlations.  
However, we have not observed such a phase in the ring model.

\underline{$F>0, G<0$}:
It follows that $\phi_{+} = \phi_{-} = 0 \mod \pi$.
In this case $\rho_{d_\alpha}$ correlations are absent, while
$j_{\perp, 1} = -j_{\perp, 2}$ show power low correlations 
at wavevector $2\pi\rho$ along the chains.
Since the currents are out of sync, this phase is less natural 
coming from the microscopic gauge theory.

Let us consider boson pairs.
First, for each $\alpha=1,2$ there are four $d_\alpha d_\alpha$ 
continuum field combinations that survive when the $\phi_\pm$ are pinned:
\begin{eqnarray}
&&\!\!\!\!\!\!\!\!\!\!\!\!
d_{\alpha P}^{(0)} d_{\alpha P}^{(\pi)}
= \eta_\alpha^{(0)} \eta_\alpha^{(\pi)} 
e^{i (\phi_\rho + e_\alpha \phi_c)}
e^{i P (\theta_\rho + e_\alpha \theta_c)}, \\
&&\!\!\!\!\!\!\!\!\!\!\!\!
d_{\alpha R}^{(k_y)} d_{\alpha L}^{(k_y)}
= e^{i [\phi_\rho + e_\alpha \phi_c
          \pm (\phi_{+} + e_\alpha \phi_{-})]} ~,
\end{eqnarray}
where in the last line the top/bottom sign corresponds to $k_y=0/\pi$.
Constructing boson pairs via $d_1 d_1 d_2 d_2$,
the field $\phi_c$ disappears, and we also set $\theta_c = 0$.
Of the sixteen combinations, six carry zero momentum and all 
have the largest scaling dimension $1/g_\rho$:
\begin{eqnarray}
&&\!\!\!\!\!\!\!\!\!\!\!\!
d_{1P}^{(0)} d_{1P}^{(\pi)} d_{2-P}^{(0)} d_{2-P}^{(\pi)}
= \eta_1^{(0)} \eta_1^{(\pi)} \eta_2^{(0)} \eta_2^{(\pi)}
e^{i 2\phi_\rho} , \\
&&\!\!\!\!\!\!\!\!\!\!\!\!
d_{1R}^{(k_y)} d_{1L}^{(k_y)} d_{2R}^{(k_y)} d_{2L}^{(k_y)}
= e^{i (2\phi_\rho \pm 2\phi_{+})} , \\
&&\!\!\!\!\!\!\!\!\!\!\!\!
d_{1R}^{(k_y)} d_{1L}^{(k_y)} d_{2R}^{(k_y+\pi)} d_{2L}^{(k_y+\pi)}
= e^{i (2\phi_\rho \pm 2\phi_{-})} .
\end{eqnarray}
The second and third lines will likely have smaller amplitudes,
since $\phi_{+}$ and $\phi_{-}$ still fluctuate a little about the 
pinned values.

We now examine the internal pairing structures.
Corresponding to the first line above, we have:
\begin{eqnarray}
\Phi_{1P(0), 1P(\pi), 2-P(0), 2-P(\pi)}(y_1, y_2, \xi)
= \\
= -4 \delta_{y_1 = y_2} 
\sin\left(\frac{k_{F1}^{(0)} - k_{F1}^{(\pi)}}{2} \xi \right) 
\sin\left(\frac{k_{F2}^{(0)} - k_{F2}^{(\pi)}}{2} \xi \right)  
\nonumber \\
-4 \delta_{y_1 \neq y_2}
\cos\left(\frac{k_{F1}^{(0)} - k_{F1}^{(\pi)}}{2} \xi \right) 
\cos\left(\frac{k_{F2}^{(0)} - k_{F2}^{(\pi)}}{2} \xi \right) ~.
\nonumber
\end{eqnarray}
The $\delta_{y_1 = y_2}$ piece describes a pair formed by bosons in
the same chain, while $\delta_{y_1 \neq y_2}$ is a pair straddling
the two chains.  When say $k_{F1}^{(0)} \approx k_{F1}^{(\pi)}$,
which becomes more accurate as we decrease $J_\perp$, the amplitudes
for the latter are significantly stronger, i.e., pairs are 
predominantly straddling the two chains.

Corresponding to the other zero-momentum combinations,
we have
\begin{eqnarray}
\Phi_{1R(k_y), 1L(k_y), 2R(k^\prime_y), 2L(k^\prime_y)}(y_1, y_2, \xi) = \\
= 4 e^{i (k_y+k^\prime_y) (y_1+y_2)} 
\sin(k_{F1}^{(k_y)} \xi) \sin(k_{F2}^{(k^\prime_y)} \xi) ~.
\nonumber
\end{eqnarray}
If $k_y = k^\prime_y$, this is independent of whether the two sites are
on the same or different chains.
On the other hand, if $k_y \neq k^\prime_y$, this has opposite signs if
the two sites are on the same versus different chains.

All of the above pair-functions are even if both particles are moved
perpendicular to the chains.  In particular, $+45^\circ$ and $-45^\circ$
diagonal pairs have the same amplitudes.  This is what we call
s-wave pairing.  Putting together all zero momentum contributions to
$b(x_1, y_1) b(x_2, y_2)$, we have
\begin{eqnarray}
e^{i 2\phi_\rho} &\Bigl[& 
2 \Gamma \Phi_{1R(0), 1R(\pi), 2L(0), 2L(\pi)} \\
&+& e^{ i 2\phi_+} \Phi_{1R(0),   1L(0),   2R(0),   2L(0)} \\
&+& e^{-i 2\phi_+} \Phi_{1R(\pi), 1L(\pi), 2R(\pi), 2L(\pi)} \\
&+& e^{ i 2\phi_-} \Phi_{1R(0),   1L(0),   2R(\pi), 2L(\pi)} \\
&+& e^{-i 2\phi_-} \Phi_{1R(\pi), 1L(\pi), 2R(0),   2L(0)} 
\Bigr] ~,
\end{eqnarray}
where $\Gamma = \eta_1^{(0)} \eta_1^{(\pi)} \eta_2^{(0)} \eta_2^{(\pi)}$
and eventually drops out since both $\Gamma \cos(2\phi_+)$ and 
$\Gamma\cos(2\phi_-)$ are determined uniquely by the signs of the
$F$ and $G$ couplings: 
$\Gamma e^{\pm i 2\phi_+} = {\rm sign}(F)$,
$\Gamma e^{\pm i 2\phi_-} = -{\rm sign}(G)$.
Note also that the contributions in the last four lines will have
smaller amplitudes since $\phi_+$ and $\phi_-$ fluctuate a little
around the pinned values.  So the numerically largest contribution
is given by the first line, while the smaller contributions
from the last four lines will add to it with relative signs that 
depend on the signs of the $F$ and $G$.

Summarizing, the DBL[2,2] is unstable towards a boson-paired phase 
with s-wave pairs carrying zero momentum.  
The resulting phase is roughly similar to the s-wave paired phase 
discussed as a possible instability of the DBL[2,1] in the 
preceding section, but some details are different.
For example, coming out of the DBL[2,1] the amplitude for rung pairs 
vanishes, while out of the DBL[2,2] the rung pairs have large amplitudes 
comparable to those of the diagonal pairs.  
The latter is more similar to what the DMRG finds in the s-wave 
paired phase in the ring model, Sec.~\ref{subsec:s-paired}.
The prediction of dominant density correlations at wavevector 
$(2\pi\rho, \pi)$ also agrees with the DMRG.
One additional prediction from the theory is that the pairing
and density power laws have exponents that are inverse of each
other.  While the DMRG estimates for the present system sizes
do not satisfy this exactly, there are likely strong finite size effects,
and we would like to revisit this with larger systems.

\section{Limiting Cases in the J-K Model}
\label{app:oddends}

\subsection{DBL[2,1] as Jordan-Wigner in a Model
with No-Double Occupancy of Rungs}

In the DBL construction Eq.~(\ref{PsiDBL}), we can view
one determinant as affecting a generalized flux attachment
or Jordan-Wigner (JW) transformation and view the other determinant 
as a ground state of the new JW fermions.
In the DBL[2,1] case, the $d_2$ determinant is composed entirely
from the bonding orbitals and becomes Eq.~(\ref{detJW});
in particular, it prevents two bosons from being on the same rung.
If we consider a hard-core boson model that prohibits double occupancy 
of rungs, then ${\rm sign}({\rm det}_2)$ is the conventional 1D chain 
JW transformation on this ladder.
As can be seen from the bottom panel in Fig.~\ref{fig:ED_results},
our ring model in the DBL[2,1] phase appears, by its own dynamics,
to strongly suppress double rung occupancy.
The optimized power of ${\rm det}_2$ in the trial wavefunction is also
relatively small, making the ${\rm sign}({\rm det}_2) |{\rm det}_2|^p$ factor 
look more like the Jordan-Wigner.  Finally, the $d_1$ determinant 
is composed of the bonding and antibonding orbitals, and its optimized 
power is relatively large though smaller than $1$, which suggests 
that the JW fermions are not far from being free despite the large 
ring exchanges and the restricted rung occupancy.

On the other hand, the DBL[2,2] would have some double occupancy of
rungs and would be an example of a non-trivial JW,
but unfortunately this phase appears to be unstable as described
in Sec.~\ref{subapp:DBL22instab}.  
The instability happens because of special kinematic conditions 
satisfied here, which allow direct Josephson coupling and locking 
between enhanced boson modes.  
However, we expect that on $N$-leg ladders with $N \geq 3$, DBL phases 
will exist which cannot be described by a conventional Jordan-Wigner 
approach.

\subsection{Solution of the K-Only Model}
\label{subapp:Konly}

The pure ring model on the 2-leg ladder can be solved exactly.
In the absence of boson hopping terms ($J=0$), the number of bosons on 
each rung, $N_{\rm rung}(x) = n(x,1) + n(x,2)$, is separately conserved.
The number of bosons in each chain, $N_{\rm ch.I}$ or $N_{\rm ch.II}$, 
is also conserved.  Any rung with $N_{\rm rung} = 0$ or $2$ effectively 
breaks the system into decoupled pieces.

Consider an isolated segment of $L$ rungs with each 
$N_{\rm rung}=1$.  The ring Hamiltonian is mapped to an XY spin chain by 
identifying boson configuration $\{n(x,1), n(x,2)\} = \{1, 0\}$ 
with spin up and $\{0, 1\}$ with spin down:
\begin{equation}
H[L] = K \sum_{x=1}^{L-1} [\tau^+(x) \tau^{-}(x+1) + \Hc] ~,
\end{equation}
where $\tau$ are the usual spin-1/2 operators.
This is readily solved by free fermions, and the ground-state energy is
\begin{eqnarray}
E_{\rm gs}[L] &=& -2 K \sum_{n=1}^{n_{\rm max}} \cos\frac{\pi n}{L+1} \\
&=& -K 
\left( \frac{ \sin\frac{\pi (2 n_{\rm max} + 1)}{2(L+1)} }
            { \sin\frac{\pi}{2(L+1)} }
- 1 \right) ~,
\end{eqnarray}
where $n_{\rm max} = L/2$ if $L$ is even (in this case 
$N_{\rm ch.I} = N_{\rm ch.II} = L/2$),
while $n_{\rm max} = (L+1)/2$ if $L$ is odd 
($N_{\rm ch.I} = N_{\rm ch.II} \pm 1$).
In either case, the ground-state energy per boson is minimized by 
making $L$ large, with the asymptotic behavior
\begin{eqnarray}
\epsilon_{\rm per boson} =
\frac{E_{gs}[L]}{L} = -\frac{2 K}{\pi} + \frac{K(1-2/\pi)}{L} +
O(\frac{1}{L^2}) ~.
\end{eqnarray}

Going back to the $K$-only model on an infinitely long ladder, 
we conclude that for arbitrary boson density $\rho < 1/2$, it is 
advantageous to phase-separate into an empty region and a half-filled 
region, since this minimizes the energy per particle
($\rho > 1/2$ can be treated by particle-hole transformation).
The phase separation arises because the ring exchanges provide
an effective attraction between particles -- see also the next section.
Note that the half-filled region itself is a highly correlated state 
of bosons.

\subsection{Bound State of Two Bosons for Large K\\
and d-Wave Paired State at Low Densities}
\label{subapp:BoundState}

Consider two bosons on an otherwise empty ladder.  
In the absence of the hoppings, $J = J_\perp = 0$, the states
\begin{eqnarray}
\frac{b(x, 1)^\dagger b(x+1, 2)^\dagger
      - b(x, 2)^\dagger b(x+1, 1)^\dagger}{\sqrt{2}} |{\rm vacuum}_b \ra 
\end{eqnarray}
are degenerate ground states with energy $-K$.
Each can be viewed as a d-wave pair sitting on a plaquette.  
Small boson hopping can be included perturbatively, and the pair 
starts to hop along the ladder with amplitude $J^2 / K$.  
In fact, inspired by similar considerations in the 1D $t-J$ model
in Ref.~\onlinecite{Chen},
the zero momentum bound state can be written exactly for the 
general $J - J_\perp - K$ model, and is created by
\begin{eqnarray}
\sum_x \sum_{n=1}^\infty \left(\frac{2 J}{K} \right)^{n-1} 
\Bigl[&&  b(x, 1)^\dagger b(x+n, 2)^\dagger \\
      &-& b(x, 2)^\dagger b(x+n, 1)^\dagger \Bigr] ~.
\end{eqnarray}
This is defined for $K > 2J$ and has energy
\begin{equation}
E_{\rm bound} = -K - \frac{4 J^2}{K} ~;
\label{Ebound}
\end{equation}
the pair ``size'' is $\xi = 1/\log[K/(2J)]$. 

Note that the bound state does not utilize any rung hopping.
Also, it competes with a free state of the two bosons where each moves 
independently and where the total energy is
\begin{equation}
E_{\rm free} = -2 (2J + J_\perp) ~,
\end{equation}
assuming $J, J_\perp \geq 0$ throughout.
As an example, for $J_\perp = 0$ the bound state has lower energy
than the state with two free bosons when $K > K_c[J_\perp = 0] = 2J$, 
while $K_c[J_\perp = 0.1 J] = 2.74 J$ and $K_c$ increases with 
increasing $J_\perp$.

For a very small density of bosons, it is then natural to propose
that for $K > K_c$ the bosons will form such tightly bound pairs 
which in turn form a Luttinger liquid.  This is the picture of
the d-wave paired phase.

As discussed in the preceding section, in the absence of the
hopping (or for very large $K$), the system at any density 
phase-separates into empty and half-filled regions.
This maximizes the ring energy per particle which attains $-2K/\pi$
and, in particular, wins over the asymptotic $-K/2$ energy per particle 
in the paired state.  
However, bosons in the phase-separated half-filled region form
a non-trivial quantum state and can hardly gain any kinetic energy
once the hopping is included.  In the boson-paired liquid, on the other
hand, the pairs can move and gain some kinetic energy, which can then 
stabilize the system against phase separation.  
As a crude estimate, we ignore the kinetic energy in the phase-separated 
case and compare $-2K/\pi$ with Eq.~(\ref{Ebound}) divided per boson
and find that the boson-paired liquid is more stable for $K < 3.8J$.
Thus, there is a sizable window over which the boson-paired liquid
is energetically preferable over the free boson liquid and is stable
against phase separation.

At very low density we also need to consider more-boson bound states 
that successively improve their ring energy but lose kinetic energy, 
and the corresponding liquids could intervene between the boson-paired 
one and the phase separation.  However, it seems likely that a 
sizable region of the boson-paired phase will remain;
of course, it also competes with other phases such as DBL,
as happens in Fig.~\ref{fig:phased2} for the filling $\rho=1/9$
studied by DMRG.

\section{Trial Wavefunctions for Boson-Paired Phases}
\label{app:PsiPaired}
Let us sketch how one can construct trial wavefunctions for 
boson-paired liquids such as the d-wave paired phase at low densities
described in the last section and in Sec.~\ref{subsec:d-paired} and 
the s-wave paired phase described in Sec.~\ref{subsec:s-paired}.
Let us denote the boson pair wavefunction as $\phi(r - r')$.
For a given pairing-up of particles we could write 
$\phi(r_1 - r_2) \phi(r_3 - r_4) \dots $ times some Jastrow-type
factor for the center of mass coordinates 
$R_{12} = (r_1 + r_2)/2, R_{34} = (r_3 + r_4)/2, \dots$,
and then symmetrize this to obtain a bosonic wavefunction:
\begin{eqnarray}
\Psi &=& {\cal S} \bigl[\phi(r_1 - r_2) \phi(r_3 - r_4) \dots \\
&\times& {\rm Jastrow}(R_{12}, R_{34}, \dots) \bigr].
\end{eqnarray}
Even if we take a simpler Jastrow which is already a symmetric
function of $\{ r_i \}$, the symmetrization of the pairing part
gives a permanent which is prohibitive in numerical calculations.

An alternative way to construct a symmetric bosonic wavefunction is
\begin{eqnarray}
\Psi &=& {\rm Pf}[g(r_1 - r_2) g(r_3 - r_4) \dots ] \\
&\times& {\rm det}[r_1, r_2, r_3, \dots] 
\times {\rm Jastrow}[r_1, r_2, r_3, \dots] ~.
\end{eqnarray}
Here pairing is realized with the help of a Pfaffian, which is a BCS 
wavefunction for spinless fermions and is specified by some
pair function $g(r-r')$, while the total bosonic symmetry is recovered 
by multiplying by a second fermionic wavefunction taken for simplicity 
to be a Slater determinant of some orbitals.
Note that for tightly bound pairs when $\phi(r-r')$ and $g(r-r')$
are very short ranged the two constructions are essentially similar 
for statistically significant configurations.
The Pfaffian and determinant are simple to work with in VMC, 
which allows one to construct trial wavefunctions also beyond the 
tightly bound limit.
Of course, both the spinless fermion pair function $g(r-r')$ in the 
Pfaffian and the orbitals in the determinant need to be judiciously 
chosen to realize the particular boson pair function $\phi(r-r')$.
We should also note that such ${\rm det} \times {\rm Pf}$ wavefunction 
does not necessarily represent a paired state of bosons, and one needs 
to measure the long-distance properties first.  For example, we found 
that for a 1D chain a product of a Slater determinant and a $p_x$-wave 
BCS wavefunction has QLRO in the boson correlator when the $p_x$-pairing
problem is at weak coupling, while the boson correlator becomes
short-range -- as is needed for the boson-paired state -- only when
the $p_x$-pairing problem is at strong coupling.

\subsection{D-Wave Paired Wavefunction}
\label{subsec:PsiDW}

Consider for illustration low-density bosons in the $J_\perp = 0$ limit.
We can construct a d-wave paired state by taking
\begin{equation}
g(x_1, y_1; x_2, y_2) = \delta_{y_1 \neq y_2} {\rm sign}(y_1 - y_2) 
e^{-|x_1 - x_2|/\xi} ~
\end{equation}
in the Pfaffian part, while combining the determinant and Jastrow 
into one factor
\begin{equation}
{\rm det} \times {\rm Jastrow} = 
\left( \prod_{i<j} \sin\frac{\pi (x_i - x_j)}{L} \right)^p ~.
\end{equation}
The power of the determinant is understood as 
$({\rm det})^p = {\rm sign}({\rm det}) |{\rm det}|^p$.
The fermion pair function is $p_y$-type and here requires that the two
particles are on different chains (which is why it is strictly 
appropriate only in the $J_\perp = 0$ limit).
For an isolated pair of bosons, this indeed gives the exact d-wave pair 
of Sec.~\ref{subapp:BoundState} when $p \to 0$.  
We use this wavefunction with two variational parameters $\xi$ and $p$ 
to connect with the DMRG results at $\rho=1/9$ in 
Sec.~\ref{subsec:d-paired}.
One could construct a similar boson-paired state by taking the Pfaffian 
factor to be a solution of a general spinless BCS problem with $p_y$ 
pairing on the ladder; 
this would allow more freedom (e.g., if we want to study 
$J_\perp \neq 0$), but has not been explored.

\subsection{S-Wave Paired Wavefunction}
\label{subapp:PsiSW}

We have also tried the ${\rm det} \times {\rm Pf}$ construction for the 
s-wave paired phase of Sec.~\ref{subsec:s-paired}.
Here we take the determinant to contain both bonding and antibonding 
orbitals, which resembles the ${\rm det}_1$ in the DBL[2,1] or DBL[2,2]
containing $d_1$ fermions with dominant hopping along the chains
(in particular, the $d_1$ bonding and antibonding bands become equally 
populated in the $J_\perp = 0$ limit).  
We take the Pfaffian to be that for a $p_x$-wave spinless BCS problem 
on the ladder, which resembles $d_2$ fermions trying to enter the 
DBL[2,2] regime but becoming paired.  There are many parameters here 
such as the $d_2$ intra- and inter-chain hoppings and $p_x$ pairings.  
We have found that the boson s-paired phase from Fig.~\ref{fig:phased1} 
and Sec.~\ref{subsec:s-paired} is reproduced roughly when the $d_2$ 
BCS problem is in the weak-pairing regime for both the bonding and 
antibonding bands and has predominant inter-chain pairing. 
Note that the above $d_1-d_2$ motivation of the ${\rm det} \times {\rm Pf}$
content is very loose and is hardly inspired by any gauge theory 
thinking but rather found by trial and error. 
We also mention that there are likely other ways to construct such 
paired boson phases.  For example, coming from 2D where paired bosons 
can form a true condensate, one natural construction to explore could be 
${\rm Pf} \times {\rm Pf}$.  
Nevertheless, the ${\rm det} \times {\rm Pf}$ that we studied gives us 
complementary confidence in the DMRG identification of the 
s-wave paired phase.


\end{document}